\documentclass[12pt]{article}
\usepackage{epsfig}
\usepackage{comment}
\usepackage{latexsym}
\usepackage{color}
\usepackage{amsmath}

\newcommand{\mysquare}[0]{\raise-.2ex\hbox{{\Large$\Box$}}}
\def\lsim{\mathrel{\rlap {\raise.5ex\hbox{$ < $}}
{\lower.5ex\hbox{$\sim$}}}}
\def\gsim{\mathrel{\rlap {\raise.5ex\hbox{$ > $}}
{\lower.5ex\hbox{$\sim$}}}} \topmargin -1.5cm \textheight=22.5cm \textwidth=16.5cm
\catcode`\@=11
\newcount\hour
\newcount\minute
\newtoks\amorpm
\hour=\time\divide\hour by60 \minute=\time{\multiply\hour by60 \global\advance\minute by-\hour}
\edef\standardtime{{\ifnum\hour<12 \global\amorpm={am}%
        \else\global\amorpm={pm}\advance\hour by-12 \fi
        \ifnum\hour=0 \hour=12 \fi
        \number\hour:\ifnum\minute<10 0\fi\number\minute\the\amorpm}}
\edef\militarytime{\number\hour:\ifnum\minute<10 0\fi\number\minute}
\def\draftlabel#1{{\@bsphack\if@filesw {\let\thepage\relax
   \xdef\@gtempa{\write\@auxout{\string
      \newlabel{#1}{{\@currentlabel}{\thepage}}}}}\@gtempa
   \if@nobreak \ifvmode\nobreak\fi\fi\fi\@esphack}
        \gdef\@eqnlabel{#1}}
\def\@eqnlabel{}
\def\@vacuum{}
\def\draftmarginnote#1{\marginpar{\raggedright\scriptsize\tt#1}}
\def\draft{\oddsidemargin -.2truein
        \def\@oddfoot{\sl preliminary draft \hfil
        \rm\thepage\hfil\sl\today\quad\militarytime}
        \let\@evenfoot\@oddfoot \overfullrule 3pt
        \let\label=\draftlabel
        \let\marginnote=\draftmarginnote
   \def\@eqnnum{(\theequation)\rlap{\k

 ern\marginparsep\tt\@eqnlabel}%
\global\let\@eqnlabel\@vacuum}  }

\newcommand{\be}[0]{\begin{equation}}
\newcommand{\ee}[0]{\end{equation}}
\newcommand{\ba}[0]{\begin{eqnarray}}
\newcommand{\ea}[0]{\end{eqnarray}}

\def\bs{\begin{subequations}}
\def\es{\end{subequations}}

\def\thebibliography#1{%
\vskip 0.5cm \centerline{\bf \Large References}
\list{%
[\arabic{enumi}]}{\settowidth\labelwidth{[#1]} \leftmargin\labelwidth \advance\leftmargin\labelsep
\usecounter{enumi}}
\def\newblock{\hskip .11em plus .33em minus .07em}
\sloppy\clubpenalty4000\widowpenalty4000 \sfcode`\.=1000\relax}

\renewcommand{\theequation}{\arabic{section}.\arabic{equation}}

\renewcommand{\section}{\setcounter{equation}{0}\@startsection
{section}{1}{0mm}{-\baselineskip}{0.5\baselineskip} {\normalfont\Large\bfseries}}

\renewcommand{\subsection}{\@startsection
{subsection}{2}{0mm}{-\baselineskip}{0.5\baselineskip} {\normalfont\large\bfseries}}

\renewcommand{\subsubsection}{\@startsection
{subsubsection}{3}{0mm}{-\baselineskip}{0.5\baselineskip} {\normalfont\normalsize\slshape}}

\usepackage{amssymb,amsfonts}
\usepackage{graphicx}
\usepackage{cite}

\newcommand{\ie}{{\em i.e. }}

\newcommand{\Z}{\mathbb{Z}}

\newcommand{\N}{{\cal N}}

\renewcommand{\O}{{\cal O}}
\renewcommand{\Re}{{\rm Re}}

\newcommand{\abs}{|}
\newcommand{\where}{\mbox{where}}
\newcommand{\with}{\mbox{with}}
\newcommand{\when}{\mbox{when}}
\renewcommand{\and}{\mbox{and}}

\newcommand{\tg}{\tilde{g}}
\newcommand{\tm}{\tilde{m}}
\newcommand{\F}{{\cal F}}
\renewcommand{\P}{{\cal P}}

\newcommand{\K}{{\cal K}}

\newcommand{\I}{{\cal I}}

\newcommand{\A}{{\cal A}}
\newcommand{\B}{{\cal B}}
\renewcommand{\F}{{\cal F}}
\newcommand{\ve}{{\varepsilon}}
\renewcommand{\max}{{\mbox{\tiny max}}}
\newcommand{\BC}{{\mbox{\tiny BC}}}

\newcommand{\cz}{c_{\mbox{\tiny $Z$}}}

\begin{document}
\begin{titlepage}
\begin{flushright}
CPHT--RR089.1208,
LPTENS--08/64, 
February 2009
\end{flushright}

\vspace{2mm}

\begin{centering}
{\bf\huge Attraction to a radiation-like era}\\
\vspace{3mm}
{\bf\huge in early superstring cosmologies$^\ast$}\\

\vspace{8mm}
 {\Large Fran\c cois Bourliot$^{1}$, Costas Kounnas$^{2}$ and Herv\'e~Partouche$^1$}

\vspace{2mm}

$^1$  {Centre de Physique Th\'eorique, Ecole Polytechnique,$^\dag$
\\
F--91128 Palaiseau cedex, France\\
{\em Francois.Bourliot@cpht.polytechnique.fr} \\
{\em Herve.Partouche@cpht.polytechnique.fr}} 

\vspace{2mm}

$^2$ Laboratoire de Physique Th\'eorique,
Ecole Normale Sup\'erieure,$^\ddag$ \\
24 rue Lhomond, F--75231 Paris cedex 05, France\\
{\em  Costas.Kounnas@lpt.ens.fr}

\vspace{5mm}

{\bf\Large Abstract}

\end{centering}
\vspace{4mm}

\noindent Starting from an initial classical four dimensional flat background of the heterotic or type II superstrings,
we are able to determine at the string one-loop level the quantum corrections to the effective potential 
due to the spontaneous breaking of supersymmetry by ``geometrical fluxes". Furthermore, considering  a gas of strings at finite temperature,  the full ``effective thermal potential" is determined, giving rise to an effective non-trivial pressure. The backreaction of the quantum and thermal corrections to the space-time metric as well as  to the moduli fields induces a cosmological evolution that depends on the early time initial conditions and the number of spontaneously broken supersymmetries. We show that for a whole set of initial conditions, the cosmological solutions converge at late times to two qualitatively different trajectories: They are either attracted to {\em (i)} a thermal evolution  similar to a radiation dominated cosmology, implemented by a coherent motion of some moduli fields, or to {\em (ii)} a  ``Big Crunch" non-thermal  cosmological evolution dominated by the non-thermal part of the effective potential or the moduli kinetic energy. During the attraction  to  the radiation-like era, periods of accelerated cosmology can occur. However, they do not give rise to enough inflation ($e$-fold $\simeq 0.2$) for the models we consider, where $\N\ge 2$ supersymmetry is spontaneously broken to $\N=0$.

\vspace{3pt} \vfill \hrule width 6.7cm \vskip.1mm{\small \small \small
  \noindent $^\ast$\ Research 
partially supported by the European ERC Advanced Grant 226371, ANR (CNRS-USAR) contract 05-BLAN-0079-02, and CNRS PICS contracts  3747 and 4172.\\ 
$^\dag$\ Unit{\'e} mixte du CNRS et de l'Ecole Polytechnique,
UMR 7644.}\\
 $^\ddag$\ Unit{\'e} mixte  du CNRS et de l'Ecole Normale Sup{\'e}rieure associ\'ee \`a
l'Universit\'e Pierre et Marie Curie (Paris
6), UMR 8549.

\end{titlepage}
\newpage
\setcounter{footnote}{0}
\renewcommand{\thefootnote}{\arabic{footnote}}
 \setlength{\baselineskip}{.7cm} \setlength{\parskip}{.2cm}

\setcounter{section}{0}


\section{Introduction}
A  sensible theoretical description of cosmology has certainly to take
into account both the quantum and thermal effects. It is naturally the
case for an early phase of the history, close to the Planck time, where quantum
fluctuations are of same order of magnitude as the whole size of the
universe and where the temperature is extremely high.
 It has to be the case as well at very late times, 
where the description of the cosmology
should involve the electroweak phase transition and provide the
missing link to the quantum particle physics of the standard model (or
its supersymmetric stringy extension), account for realistic large scale
structure formations, as well as describe correct temperature properties 
of the cosmic microwave background. 

In the framework of strings, the theory of quantum gravity is well defined \cite{GSW}
(at least in certain cases)  and can be  considered at finite temperature as well\cite{AtickWitten,RostKounnas,AKADK}. 
Thus, in  the stringy framework it is natural to attempt to describe the cosmological
evolution of our universe. In the very early times and when the temperature
approaches the so called  Hagedorn temperature 
\cite{Hagedorn,AtickWitten,RostKounnas,AKADK}, we are faced to non-trivial 
stringy singularities indicating a high temperature non-trivial 
phase transition.
In the literature, there are many speculative proposals concerning the nature of this 
transition \cite{AtickWitten,AKADK,GV,BV,GravFluxes}.

 Conceptually, it is quite exotic to understand the early stringy phase utilizing 
the standard field theoretical geometrical notions  \cite{CosmoTopologyChange,KTT,MSDS,GravFluxes}.
Indeed, following for instance Refs \cite{CosmoTopologyChange,MSDS}, the notions of geometry and even topology break down in this very early-time era. Only by considering stringy approaches based on 2-dimensional conformal field theory techniques, there is a hope to analyze further this phase.
Actually, some conceptual progress in this direction has been achieved recently\cite{MSDS}, but we are not yet in a position to give quantitative descriptions of our stringy universe at these very early times. This stringy-conceptual obstruction however is not an obstacle to the study of the cosmological 
evolution of our universe for much later times than the Hagedorn transition era, namely for late times 
such that $t\gg t_H$, \cite{Cosmo-0, Cosmo-1,Cosmo-2}.
  
 A way to bypass the Hagedorn transition ambiguities consists in assuming the emergence 
 of three large space-like directions for $t\gg t_H$,  describing the three dimensional space of our universe, and  possibly some internal space directions of an intermediate size, characterizing the scale of the spontaneous  breaking of supersymmetry, via geometrical fluxes, or even other (gauge or else) symmetry breaking scales  \cite{BEKP}.  Within this assumptions, the  ambiguities of the ``Hagedorn transition exit  at $t_E$"  can be parametrized,
 for $t\ge t_E$,  in terms of initial  boundary condition data at  $t_E$. This is precisely the scope of the present work, were we  focus on the intermediate cosmological era:  $ t_E\le t \le t_w$, namely, after the  ``Hagedorn transition exit"  and before the electroweak symmetry breaking phase transition at $t_w$. 
 
 Modulo the initial  boundary condition data at  $t_E$, the stringy description of the  intermediate cosmological era, turns out to be under control, at least for the string backgrounds where supersymmetry is spontaneously broken by ``geometrical fluxes" \cite{GeoFluxes,OpenFluxes}. These fluxes are implemented by a ``stringy  Scherk-Schwarz mechanism" involving  $n$-internal compactified dimensions which are  coupled non-trivially to some of the R-(super-)symmetric charges \cite{SS,Rohm,KouPor} \cite{Cosmo-1,Cosmo-2,GravFluxes}. 
Furthermore,  the  finite temperature effects are also implemented by a stringy Scherk-Schwarz compactification attached to the  Euclidean time circle of radius $\beta=1/T$\cite{AtickWitten,RostKounnas,AKADK,Cosmo-1,Cosmo-2,GravFluxes}. Here, the R-symmetry charge is just the helicity, so that space-time bosons are periodic while fermions are anti-periodic along the $S^1$ Euclidean time circle.

The breaking of supersymmetry and the quantum canonical ensemble at temperature
$T$ give rise to a non-trivial free energy at the one-loop string level. The latter is evaluated in a large regularized 3-space volume $V_3$. This amounts to considering the free energy of a gas of approximately free strings, 
\be
F=-\frac{\ln Z}{\beta}\simeq - {Z^{\mbox{\tiny (string)}}_{\mbox{\tiny $1$-loop}}\over \beta}\, , 
\ee
where $Z={\rm Tr}\, e^{-\beta H}$ is the partition function at finite temperature and  $Z^{\mbox{\tiny (string)}}_{\mbox{\tiny $1$-loop}}$ is the one-loop connected graph \ie  genus one vacuum to vacuum string amplitude computed in the Euclidean background. In these equalities, we use the fact that the free energy of an infinite number of degrees of freedom can be found by a single string loop computation \cite{Ditsas:1988pm} \cite{Cosmo-0,Cosmo-1,Cosmo-2}. The pressure derived from the free energy, $P=-\partial F/\partial V_3$, has been computed for heterotic and type II models\footnote{Type I models realized as orientifolds of type IIB can also be considered. They can be dual to heterotic ones with an initially $\N=4$, $2$ or even $1$ supersymmetry \cite{het-TIdual}. }. Supersymmetry is spontaneously broken by introducing geometrical fluxes \textit{\`a la} Scherk-Schwarz involving $n$ internal directions and by finite temperature effects. The pressure acts as a source occurring at 1-loop that backreacts on the original classical space-time metric and moduli fields background. In some cases, the perturbed equations of motion for the
metric and moduli are shown to admit a cosmological evolution
describing a radiation-like dominated era\cite{AKradiation,Cosmo-0,Cosmo-1,Cosmo-2}. The latter is characterized by a time
evolution where the temperature $T(t)$, the supersymmetry breaking scale $M(t)$
and the inverse scale factor $1/a(t)$ remain proportional\cite{AKradiation,Cosmo-0,Cosmo-1,Cosmo-2}: 
\be
\mbox{\em Radiation Dominated Solution :} ~~~T(t) \propto M(t) \propto 1/a(t) ~~~{\rm and }~~~H^2 \equiv \left( {\dot a\over a} \right)^2\propto {1\over a^4}\,.~~~~~~
\ee
Very different time evolutions also  exist when temperature effects are neglected  \cite{Cosmo-2}. In this case, the supersymmetry breaking scale evolves like:
  \be
\mbox{\em Moduli Dominated Solution :} ~  M(t)\propto 1/a(t)^{2+4/n}~~ {\rm with}~~T(t)\ll M(t)~~{\rm and }~~H^2 \propto {1\over a^{8(n+2)/n}}\,.
\ee

In this work we would like to examine the late time evolution of the universe in the intermediate cosmological era,  
$ t_E\le t \le t_w$, in terms of initial boundary conditions (IBC) after the ``Hagedorn transition exit  at $t_E$".
We show that the Radiation Dominated Solution (RDS) and the Moduli Dominated Solution (MDS) obtained for particular IBC are actually generic. For instance, for $n=1$,  there is a whole basin of IBC such that the resulting cosmological evolutions converge -- depending on the model --  either to the RDS or to the MDS. The first attractor corresponds to an expanding evolution controlled by a Friedmann equation $H^2\propto 1/a^4$, where $H$ is the Hubble parameter.  The second one describes a Big Crunch behavior dominated by the non-thermal radiative corrections. In addition, all models admit a second basin of IBC that yields  to another Big Crunch behavior, dominated by the moduli kinetic energy \ie where $H^2\propto 1/a^6$.  Within the scope of our present work, the Big Crunch evolutions can be trusted as long as Hagedorn like singularities are not reached. Our analysis is also limited to the cases where the initial supersymmetry is higher or equal to two, $\N\ge 2$. Although the $\N=1$ case is quite similar in some respects, it is much more involved in issues like the appearance of an effective cosmological term. It  will be analyzed elsewhere.

Restricting to cases with $\N \ge 2$ initial supersymmetry, the attraction to the radiation era depends on the IBC: The convergence to the RDS is either exponential or via damping oscillations. During this process, periods of accelerated expansion can occur. However, they are characterized by  an $e$-fold number of order $0.2$, which is too low to account for realistic astrophysical observations. It would be very interesting to extend our analysis to models with $\N=1$ initial supersymmetry and see if the $e$-fold number is larger enough for an inflation era to occur. 

In Section 2, the case of a supersymmetry breaking using a single internal
direction (\ie $n=1$) is studied in details. We classify the models and find local basins of attraction. This analysis is completed by a more general exploration of the set of IBC by numerical simulations. In Section 3, the attraction mechanism to the RDS is extended to models where $n=2$ internal directions participate in the supersymmetry breaking. We determine in Section 4 the $e$-fold number of the periods of acceleration. Our results, their limitations and our perspectives are summarized in Section 5.    


\section{Susy breaking involving {\em n} = 1  internal dimension  }
\label{one mod}

In this section, we summarize first the results and notations
introduced in Ref. \cite{Cosmo-1} and then we analyze the 
attractor mechanism. 
The starting background  can be  the
heterotic, type I or type II superstring compactified on a 6-dimensional space ${\cal
  M}$ that can be either $S^1(R_4) \times T^5$ or $S^1(R_4) \times S^1\times K3$. The sizes of ${T^5}$ and $S^1\times K3$ are  chosen to be of order the Planck length\footnote{We relax this hypothesis in Ref. \cite{BEKP}.}. 
The supersymmetry breaking is introduced by ``geometrical fluxes" via the coupling of  the momentum and winding numbers  
of the Scherk-Schwarz circle $S^1(R_4)$ to an R-symmetry charge \cite{SS,Rohm,KouPor,RostKounnas,Cosmo-1,Cosmo-2}. 
The radius of the circle  $R_4$ controls the supersymmetry breaking scale $M$. In all applications, we will replace $K3$ by its orbifold limit, $K3 \sim T^4/\Z_2$.  

 In the heterotic and type I cases initially $\N=4$ or  $\N=2$, the geometrical fluxes break
spontaneously all the supersymmetries, $\N \to 0$. In type II, when the  $S^1(R_4)$ coupling involves non-trivially left- and right- moving R-symmetry charges, all initial $\N=8,$ $\N=4,$ or $\N=2 $ supersymmetries are spontaneously broken. When the $S^1(R_4)$ coupling is left- right- ``asymmetric''  \ie involves non-trivially a left- (or right-) moving R-symmetry charge only, half of the supersymmetries still survive: $\N\to \N/2$.

\subsection{General setup, gravity-moduli equations}
In all cases the computation of the free energy involves for the heterotic or type II superstring the
background:  
\be
S^1(R_0)\times T^3\times S^1(R_4)\times {\cal M}_5\, ~~{\rm with~either}~~{\cal M}_5 = T^5~ {\rm or}~~{\cal M}_5 =S^1\times T^4/ \Z_2\, .
\ee
The $T^3$ factor stands for the external space with regularized volume $V_3$ 
sent to infinity once the thermodynamical limit is taken. $S^1(R_0)$
is the Euclidean time circle of period $\beta=2\pi R_0$ that defines the
inverse temperature (in the string frame). In this direction, the Scherk-Schwarz circle
amounts to coupling the momentum and winding numbers of the Euclidean time circle
to the space-time fermion number \cite{AtickWitten,RostKounnas,AKADK,Cosmo-1,Cosmo-2,GravFluxes}. The computation of the free energy density is
well defined if the moduli $R_0$ and $R_4$ responsible for
the spontaneous supersymmetry breaking are larger than the Hagedorn
radius $R_H$, which is of order 1 in string units. At the Hagedorn
radius, some winding modes become tachyonic so that the free energy
develops a severe IR divergence\cite{AtickWitten,RostKounnas,AKADK,Cosmo-1,Cosmo-2,GravFluxes}. 
If our goal is not to describe the phase
transition that is occurring at the Hagedorn radius, we can suppose
that in the intermediate cosmological era, $R_0$ and $R_4$ are large. 
In Refs \cite{Cosmo-1,Cosmo-2}, it is  shown  that up to exponentially
 small terms of order $e^{-\pi R_0}$ or $e^{-\pi R_4}$, the pressure derived from 
 the free energy density takes the following form\cite{AKradiation,Cosmo-0,Cosmo-1,Cosmo-2},  
\be
\label{P1mod}
P=T^4\, p(z)\qquad \where \qquad e^z={M\over T}\, ,
\ee
with the temperature $T$ and supersymmetry breaking scale $M$ defined as,
\be
T={1\over 2\pi R_0\sqrt{\Re S}}\; , \qquad M={1\over 2\pi R_4\sqrt{\Re S}}\, . 
\ee
In these expressions, the  factors involving the dilaton in 4
dimensions, $\Re S=e^{-2\phi_D}$, arise when we work in Einstein
frame. In (\ref{P1mod}), $p$ is a linear sum of functions
of $z$ with integer coefficients $n_T$ and $n_V$ \cite{Cosmo-1,Cosmo-2}, 
\be
\label{press}
p(z)=n_T f(z)+n_V \tilde f(z)\, ,
\ee
where
\be
f(z)={\Gamma(5/2)\over \pi^{5/2}}\sum_{m_0,m_4}{e^{4z}\over
  [(2m_0+1)^2 e^{2z}+(2m_4)^2]^{5/2}}\; ,\qquad \tilde
f(z)=e^{3z}f(-z) \, , 
\ee 
and the sums are over the integers in $\Z$. $n_T$ is the number of massless
states of the model. If the Scherk-Schwarz  circle $S^1(R_4)$ acts
non-trivially on the left- and right- moving R-charges, $n_V$ satisfies the inequality:
\be
-1\le{n_V\over n_T}\le1\, ,
\ee
and depends on the choice of R-symmetries used to spontaneously break 
all the supersymmetries \cite{Cosmo-1,Cosmo-2}. For an asymmetric $S^1(R_4)$ coupling
\ie trivial on the left- (or right-) moving R-symmetry charges, 
the supersymmetry breaking is partial and thus,
there is no contribution $\tilde f$ to $p$. In that case, $M$ is the mass of the 
$\N/2$ massive gravitini. In the following, we treat both kinds of 
models keeping in mind that $n_V=0$ for asymmetric R-symmetry coupling to
 the momenta and windings of the 4$^{\mbox{\footnotesize th}}$ internal direction. 

From a low energy effective field theory point of view, the pressure
$P$  is minus the effective potential at finite temperature. It is a
source for the quantum fields, namely the dilaton $\phi_D$ and the
modulus $R_4$, coupled to 
gravity\cite{AKradiation,Cosmo-0,Cosmo-1,Cosmo-2}. It also introduces the ``thermal
field'' $R_0$ to which there is no associated quantum fluctuation, as
can be seen from the string model in light-cone gauge where oscillator
modes in the direction 0 are gauged away. The source $P$ is a 1-loop
correction that backreacts on the classical 4-dimensional Lorentzian
background, via the effective field theory action, 
 \be
 \label{action}
S = \int d^4x \sqrt{-g}\left[ {1\over 2}\,
 R-g^{\mu\nu}\left(\partial_\mu \phi_D\, \partial_\nu\phi_D+
 {1\over2}\, {\partial_\mu R_4\, \partial_\nu R_4\over R_4^2}\right)+P \right]\, , 
\ee
where we write explicitly the fields with non-trivial backgrounds only. Note that
computing the 1-loop correction to the kinetic terms is not needed. Indeed,
the wave function renormalization of the moduli, (to recast the kinetic terms at 1-loop in
a canonical form), would  introduce an additional correction to the
pressure at second order only. Thus, the full low energy dynamics at first order in string
perturbation theory is described by the above action, (\ref{action}). 

Before writing the equations of motion,  it is useful  to redefine the fields, 
\be 
\phi:= \sqrt{2\over 3}\, (\phi_D-\ln R_4)\; , \qquad \phi_\bot := {1\over
  \sqrt{3}}\, (2\phi_D+\ln R_4)\, ,  
\ee
so that the action becomes:
 \be
S = \int d^4x \sqrt{-g}\left[ {1\over 2}\, R-{1\over 2}\left((\partial
    \phi)^2+(\partial \phi_\bot )^2\right)+P \right]\, ,  
\ee
where the pressure and the supersymmetry breaking scale $M$ take the form,
\be
\label{M4}
P=M^4\, e^{-4z}p(z)\; ,\qquad e^z={M\over T}\; , \qquad
M={e^{\alpha \phi}\over 2\pi}\qquad \with \quad \alpha=\sqrt{3\over
  2}\, . 
\ee
This shows that in  the 2-dimensional moduli space $(\phi_D, R_4)$,  
the effective potential lifts the ``no-scale modulus'' $\phi$, while keeping flat the orthogonal direction $\phi_\bot$. 
Assuming an homogeneous and isotropic cosmological metric with a flat  3-dimensional subspace, 
\be
\label{FRW}
ds^2=-N(t)^2dt^2+a(t)^2\left(dx_1^2+dx_2^2+dx_3^2\right) ,
\ee 
the Friedmann equation in the $N(t)=1$ gauge takes the form,
\be
\label{eq1}
3H^2={1\over 2} \dot\phi^2+{1\over 2} \dot \phi_\bot^2+\rho\, , 
\ee
where the energy density $\rho$ is defined by the identity,
\be
\label{rho}
\rho+P=T\, {\partial P\over \partial T}\, ,
\ee
and $P$ is understood as a function of the independent variables $T$ and $M$. 
As shown  in Ref.\cite{Cosmo-2}, the expression of $\rho$ follows
from the variational principle applied to a  gauge choice of the function
$N=1/T$. It is fundamental to stress  here that the energy density defined by the variational principle 
is identical to the one derived from the second law of thermodynamics. From Eqs  (\ref{P1mod}) and
(\ref{rho}),  it follows that: 
\be
\label{r}
\rho=T^4\, r(z) \quad \where\quad  r=3p-p_z\, ,
\ee
where  $z$-indices stand for derivatives with respect to $z=\ln (M/T)$.   Varying
the action with 
respect to the scale factor $a$ or the scalar fields $\phi$ and
$\phi_\bot$, we obtain: 
\begin{eqnarray}
&&\label{eq2}
2\dot H+3H^2=-{1\over 2} \dot \phi^2-{1\over 2} \dot \phi_\bot^2-P\, ,\\
&&\label{eq3}
\ddot{\phi}+3H\dot\phi={\partial P\over \partial \phi}\equiv \alpha (3P-\rho)\, ,\\
&&\label{eq4}
\ddot\phi_\bot+3H\dot\phi_\bot=0\qquad \Longrightarrow\qquad \dot
\phi_\bot=\sqrt{2}\; {c_\bot\over a^3}\, ,
\end{eqnarray}
where $c_\bot$ is an arbitrary integration constant. Note that the
last identity in Eq. (\ref{eq3}) is a consequence of the fact that $P$
is a dimension four object, $(M\partial_M+T\partial_T)P\equiv 4P$, and from
the definitions (\ref{M4}) and (\ref{rho}).  The conservation of the total  energy-momentum tensor
follows from  the Einstein equations (\ref{eq1}) and (\ref{eq2}):
\be
\label{cE}
{d\over dt} 
\left({1\over 2} \dot \phi^2+{1\over 2} \dot
  \phi_\bot^2+\rho\right)+
3H\left(\dot\phi^2+\dot\phi_\bot^2+\rho+P\right)=0\, .
\ee
Using further the equations of the fields $\phi$ and
$\phi_\bot$,  we obtain from Eq. (\ref{cE}) that the $\{\rho,P\}$-system 
is not isolated but is coupled non-trivially
to the ``no-scale modulus field", $\phi$ \cite{Noscale,StringyNoscale}:
\be
\label{cEbis}
\dot\rho+3H(\rho+P)+\alpha \dot\phi(3P-\rho)=0\, .
\ee


\subsection{Classification of the models, effective potential approach}
\label{classi}

Our aim is to analyze the time behavior of the system described  above. For this purpose, one can choose four independent non-linear equations, for instance:  Eqs (\ref{eq1}), (\ref{eq3}),
(\ref{eq4}) and (\ref{cEbis}), any other choice being equivalent. Furthermore, we find convenient to express the 
dependence in time in terms of a dependence in the logarithm of scale factor $a(t)$:
\be
\label{new t}
\lambda :=\ln a(t)\qquad \Longrightarrow\qquad \dot y\equiv {dy\over
  dt}=H{dy\over d\lambda}\equiv H\overset{\circ}{y}\, , 
\ee
for any function $y$. 
In particular, the definition $z=\alpha \phi - \ln(2\pi T)$ implies
$\dot T/T=H(\alpha \overset{\circ}{\phi}-\overset{\circ}{z})$ that can
be used to write 
\be
\dot \rho = HT^4\left(
  4r\alpha\overset{\circ}{\phi}+(r_z-4r)\overset{\circ}{z}\right)\, . 
\ee
This equality  is useful to express  the conservation of energy-momentum
(\ref{cEbis}) into a relation giving $\overset{\circ}{\phi}$ in terms of  $z$ and $\overset{\circ}{z}$,
\be
\label{phirond}
\alpha \overset{\circ}{\phi}  = \A_z(z)\overset{\circ}{z}-1\qquad
\where \qquad \A_z(z)={4r-r_z\over 3(r+p)}\, .
\ee
Integrating, one obtains
\be
\label{Ma}
M={e^{\A(z)}\over a}\, ,
\ee
where $\A(z)$ involves an integration constant.
Using the above  relations, the Friedmann Eq. (\ref{eq1}) becomes
\be
\label{Friedrond}
3H^2 = {3\over 8}\, H^2 \left( (\A_z(z)
\overset{\circ}{z}-1)^2-1\right)+{9\over 8}\, {e^{4(\A(z)-z)}\over
a^4}\, r(z)+{9\over 8}\, {c_\bot^2\over a^6}\, ,
\ee
or, 
\be
\label{eqh}
H^2=T^4\, {r(z) \over 3-\K(z,\overset{\circ}{z},\overset{\circ}{\phi}_\bot)}\qquad {\rm with}
\qquad \K={1\over 2\alpha^2}\left(\A_z(z)\overset{\circ}{z}-1\right)^2+{1\over 2}\, \overset{\circ}{\phi}_\bot\!\!\!{}^2\, . 
\ee  

The equations for $\phi$ and $\phi_\bot$, 
(\ref{eq3}) and  (\ref{eq4}), take the form, 
\be
\label{syseq}
\left\{
\begin{array}{l}
\displaystyle
{r(z)\over 3-\K(z,\overset{\circ}{z},\overset{\circ}{\phi}_\bot)}\left(\A_z(z)\,
  \overset{\circ\circ}{z}+\A_{zz}(z)\, \overset{\circ}z
  \!\!\!\phantom{x}^2\right)+{r(z)-p(z)\over 2}\,  \A_z(z)\,
\overset{\circ}{z}+V_z(z)=0\\ \\ 
\displaystyle {r(z)\over 3-\K(z,\overset{\circ}{z},\overset{\circ}{\phi}_\bot)}\,
\overset{\circ\circ}{\phi}_\bot +{r(z)-p(z)\over 2}\,
\overset{\circ}{\phi}_\bot =0 
\end{array}
\right. 
\ee
For the derivation of the above equations, we used the relation $~\ddot y=\dot
H\overset{\circ}{y}+H^2\overset{\circ\circ}{y}~$ for $y=\phi$ or
$\phi_\bot$ as well as the Eqs (\ref{phirond}), (\ref{eqh}) and the linear sum of (\ref{eq1}) and (\ref{eq2}). As in Ref. \cite{AKradiation}, we have introduced the notion of an effective potential $V(z)$,  defined by its derivative (and using $\alpha^2=3/2$):
\be
\label{r4p}
V_z(z)=r(z)-4p(z)\, .
\ee

We are going to see in the following section that any solution of the system  with constant $z$,  \ie  $z\equiv z_c$ where $z_c$ is an  extremum of $V(z)$,  is giving rise to a particular RDS.  The existence of such a critical point depends drastically  on the shape of the potential $V(z)$, which is determined by its asymptotic behaviors  for $z\to \pm \infty$. It is obvious  from Eq.  (\ref{press}) and the definition (\ref{rho}) of $\rho$  that $V(z)$
depends only on two parameters, namely, $n_T$ and $n_V$. The asymptotic behavior   $z\to \pm \infty$  can be easily derived:

$\bullet$ {\bf ({\em n}$_{\! \mbox{\scriptsize \bf \em V}}$/{\em n}$_{\! \mbox{\scriptsize \bf \em T}}$) $\neq$ 0,}
\be
\label{V+inf}
V(z)\underset{z\to +\infty}{\sim} - e^{4z}\,  \left({n_V\over n_T}\right)\, \left( {5 \over 4}\, n_T c_5^o
\right)\, ,~~~ ~~{\rm with}\qquad ~~c_5^o={\Gamma (5/2)\over \pi^{5/2}}\, \sum_m
{1\over |2m+1|^5}\, ,
\ee

\begin{figure}[h!]
\begin{center}
\vspace{.3cm}
\includegraphics[height=3.5cm]{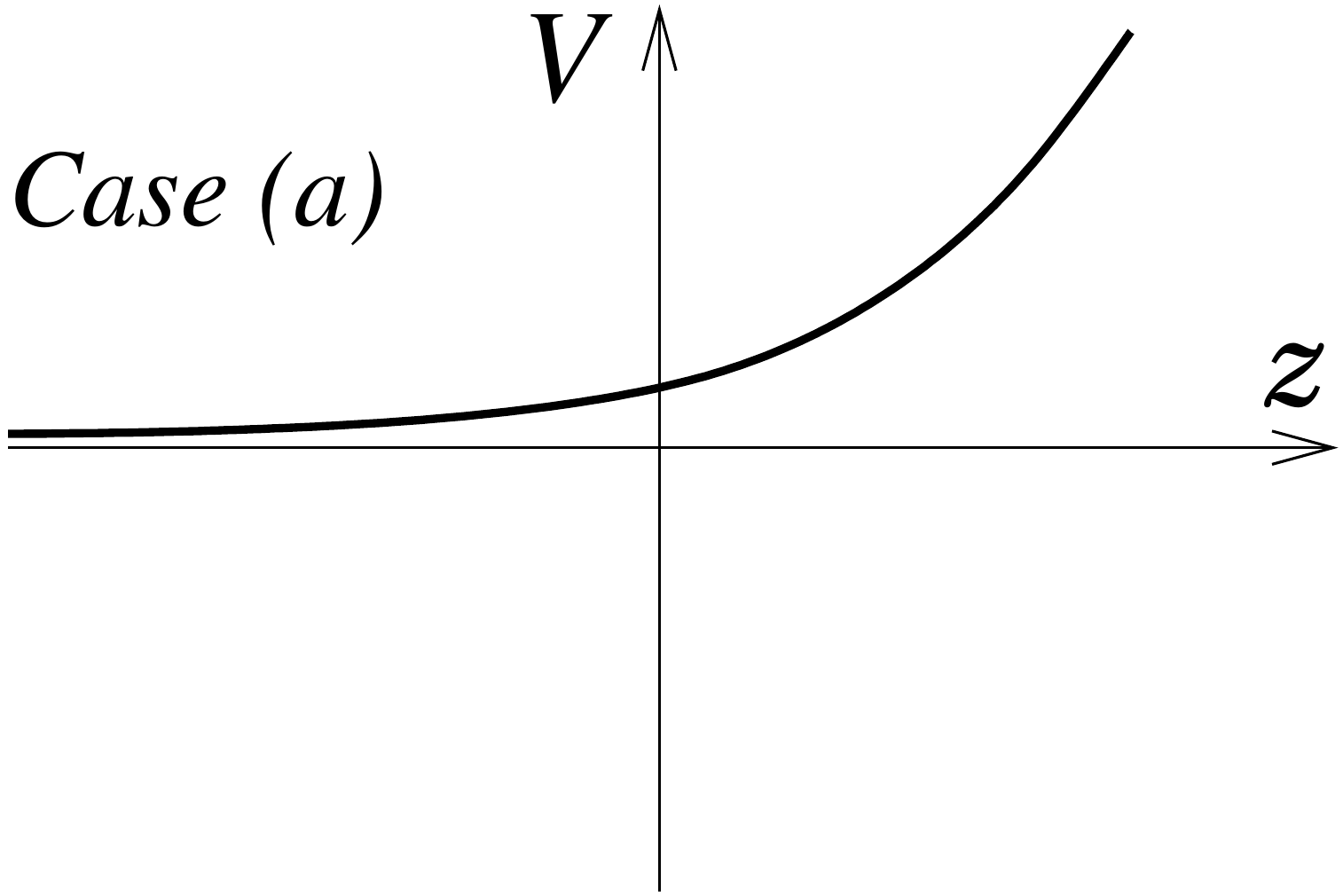}
\includegraphics[height=3.5cm]{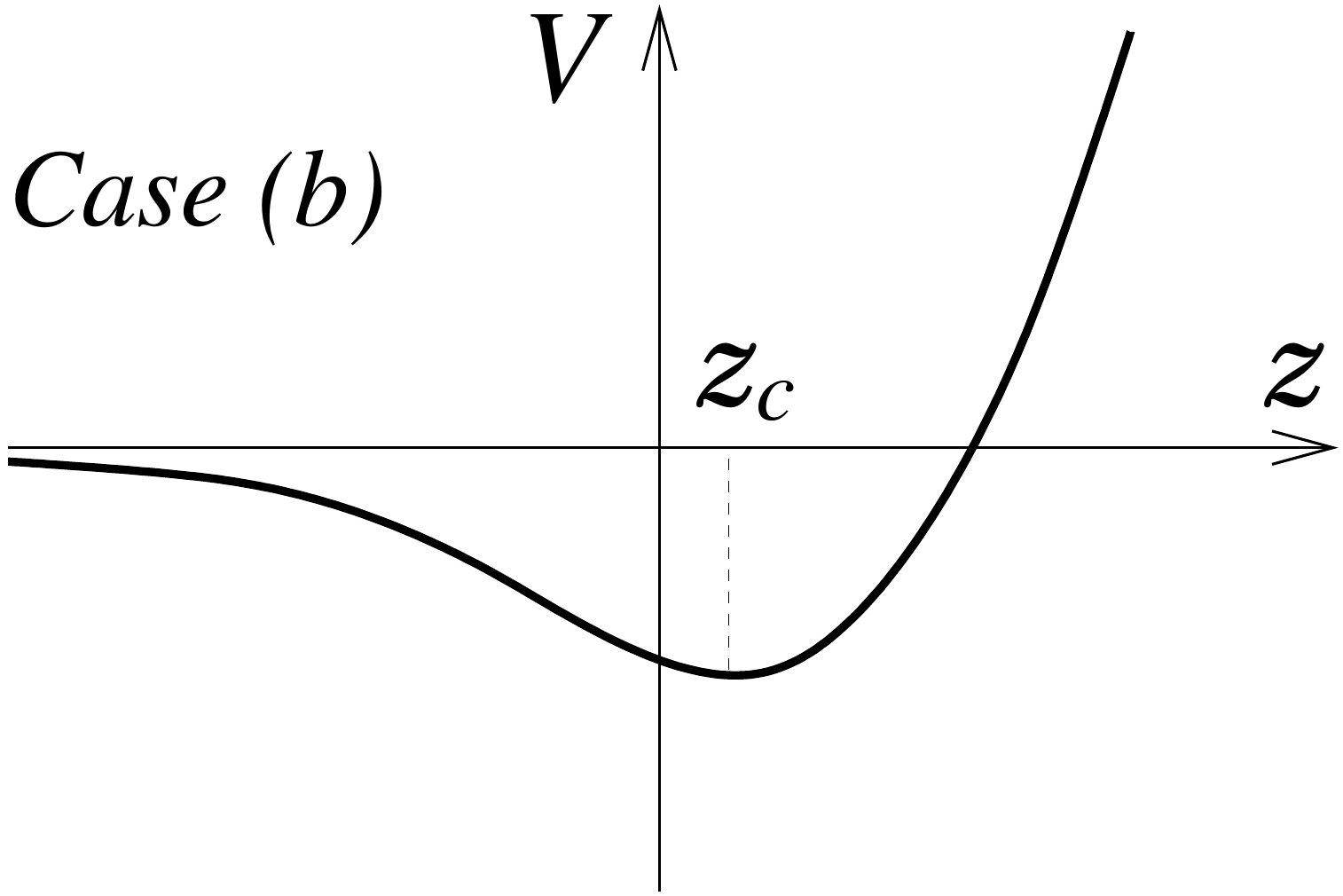}
\includegraphics[height=3.5cm]{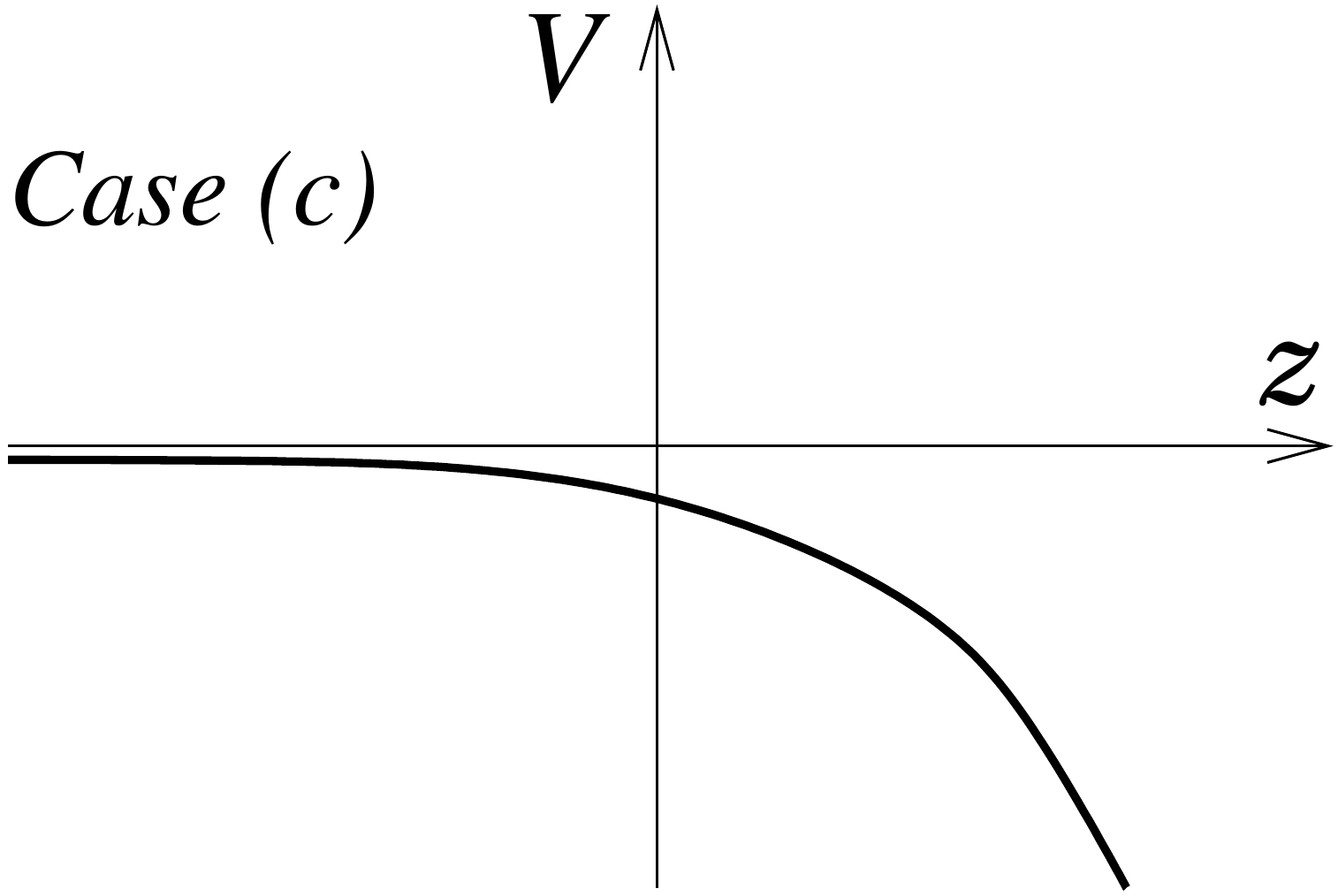}
\caption{\footnotesize \em The shape of the potential $V(z)$ in Eq. (\ref{syseq}) depends on the ratio $n_V/n_T$. The Cases (a), (b) and (c) correspond to $-1\le n_V/n_T\le -1/15$, $-1/15< n_V/n_T<0$ and $0\le n_V/n_T\le 1$, respectively.}  
\vspace{-.4cm}
\label{fig_V}
\end{center}
\end{figure}

$\bullet$ {\bf({\em n}$_{\! \mbox{\scriptsize \bf \em V}}$/{\em n}$_{\! \mbox{\scriptsize \bf \em T}}$) $\neq$ -1/15}  $\equiv -\left({\sum}_m' {1\over (2m)^4}\right)\left/
  \left({\sum_m} {1\over (2m+1)^4}\right.\right)${\bf ,}
\be
\label{V-inf}
~~V(z) \underset{z\to -\infty}{\sim} - e^{3z} \left({n_V\over
    n_T}+{1\over 15}\right)  \left({4\over 3}\, n_Tc_4^o\right)\, , ~~ {\rm with}~~ c_4^o=
{\Gamma(5/2)\over \pi^{5/2}}\, {2\over
  3}\sum_m {1\over (2m+1)^4} \, . 
\ee
According to the ratio $n_V/n_T$,  three distinct cases can actually arise, (see Fig. \ref{fig_V}),
\begin{itemize}
\item {\bf \em \large Case (a) : }   {\bf -1} $\leq$ {\bf({\em n}$_{\! \mbox{\scriptsize \bf \em V}}$/{\em n}$_{\! \mbox{\scriptsize \bf \em T}}$)} $\leq$ {\bf -1/15}\\
$V(z)$
increases monotonically  from 0 to $+\infty$. 
\item {\bf \em \large Case (b) : } {\bf -1/15} $<$ {\bf({\em n}$_{\! \mbox{\scriptsize \bf \em V}}$/{\em n}$_{\! \mbox{\scriptsize \bf \em T}}$)} $<$ {\bf 0}\\
$V(z)$ admits a unique minimum $z_c$, with $p(z_c)$ positive. When $(n_V/n_T)\to 0_-$ the critical value $z_c$ goes to $+\infty$; when  $(n_V/n_T)\to (-1/15)_+$ the critical value  $z_c\to -\infty$.
\item {\bf \em \large Case (c) : }  {\bf 0} $\leq$ {\bf({\em n}$_{\! \mbox{\scriptsize \bf \em V}}$/{\em n}$_{\! \mbox{\scriptsize \bf \em T}}$)} $\leq$ {\bf 1}\\ 
$V(z)$
decreases monotonically  from 0 to $-\infty$. 
\end{itemize}
Qualitatively, one can expect  that the system represented by  $z$ can slide along its potential. It may run away in {\em Case (a)}, $z(t)\to -\infty$, be stabilized in {\em Case (b)}, $z(t)\to z_c$, and run away in {\em Case (c)}, $z(t)\to +\infty$. Moreover, these behaviors should imply new ones by time reversal. In particular, other run away behaviors are expected: $z(t) \to +\infty$ in {\em Case (a)}, $z(t) \to \pm\infty$ in {\em Case (b)}, and $z(t) \to -\infty$ in {\em Case (c)}. However, the study of the run away $z\to-\infty$ is out  of the scope of the present work. This is due to the fact that this limit amounts to $R_4\gg R_0>1$, so that the thermal system should be studied in 5 dimensions. In the following, we will only consider the other behaviors, namely the stabilization of $z$ and the run away $z\to+\infty$. Since the latter amounts to $R_0\gg R_4>1$, it is consistent with our analysis in four dimensions but with negligible thermal effects.


\subsection{Local attraction to the RDS}
\label{constantz}

Clearly, the first equation in (\ref{syseq}) admits a constant $z$
solution whenever $V$ has an extremum, \ie in {\em Case (b)}. It
follows from Eq. (\ref{Ma}) that $M$, $T$ and $1/a$ are propotionnal during the
evolution,
\be
\label{prop}
M(t)=e^{z_c} \, T(t)= {e^{\A(z_c)}\over a(t)}\, .
\ee
The time dependence of the scale factor is dictated by the Friedmann
Eq. (\ref{Friedrond}) that simplifies to:
\be
\label{cosmoT}
3H^2={c^2_r\over a^4}+{c^2_m\over a^6}\qquad \where \qquad
c^2_r={9\over 2}\, e^{4(\A(z_c)-z_c)}\, p(z_c)\; , \quad
c^2_m={9\over 8}\, c^2_\bot\, . 
\ee
Integrating, $t$ can be expressed in terms of
the scale factor, 
\be
\label{expan}
t(a)=t_0 \int_0^{a/a_0}{x^2dx\over \sqrt{1+x^2}}\; , \quad
t_0=\sqrt{3}\,{c^2_m\over |c_r|^3}\; , \quad a_0=\left|c_m\over c_r\right|\; , \quad
\forall a\ge0\, , 
\ee
up to an overall sign. There are thus two monotonic solutions mapped to one another by time reversal. For the expanding one, the universe life starts in a ``moduli kinetic energy dominated era"
characterized by a total energy density $\simeq c^2_m/a^6$, followed by a
``radiation dominated era" with total energy density $\simeq
c^2_r/a^4$. 
For large enough times, the evolution always enters in the radiation
era: It is asymptotic to the critical solution $z\equiv
z_c$, $c_m=0$ \ie (\ref{prop}) where   
\be
\label{sol cri}
a(t)=\sqrt{t}\times  \left({4\over
    3}\, c^2_r\right)^{1/4} \; , \qquad
\dot \phi_\bot=0\, .  
\ee

To study the stability of this solution, we analyze its
behavior under small perturbations. In terms of the variable $\lambda=\ln a$,
we define, 
\be
z(\lambda)= z_c+\ve(\lambda)\; , \qquad \overset{\circ}{\phi}_\bot(\lambda)=
\overset{\circ}{\ve}_\bot(\lambda)\, , 
\ee
for small $\ve$ and $\overset{\circ}{\ve}_\bot$. At first order, the
system (\ref{syseq}) becomes 
\be
\label{line}
\left\{
\begin{array}{l}
\displaystyle \overset{\circ\circ}{\ve}+\overset{\circ}{\ve}+\xi\,
\ve=0\qquad \where\qquad \xi=10\left.{p-p_{zz}\over
    19p+p_{zz}}\right|_{z_c}\, ,\\ 
\displaystyle \overset{\circ\circ}{\ve}_\bot+\overset{\circ}{\ve}_\bot=0\, .
\end{array}
\right.
\ee
The coefficient $\xi$ is actually a function of the model dependent
parameter $n_V/n_T$, since the latter appears explicitly in the
definition of $p$ and implicitly via $z_c$. Numerically, one finds
that $\xi$ is always positive. It increases from $0$ to $+\infty$ when
$n_V/n_T$ varies from $-1/15$ to $0$.  The solution of the system
(\ref{line}) is 
\be
\label{solline}
\overset{\circ}{\ve}_\bot(\lambda)=c_3\, e^{-\lambda}\; , \; \ve(\lambda)=\left\{
\begin{array}{ll}
c_1\, e^{-\lambda(1+\sqrt{1-4\xi})/2}+ c_2\,
e^{-\lambda(1-\sqrt{1-4\xi})/2}&\mbox{if}\; 0<\xi<1/4\\ \\ 
e^{-\lambda/2}(c_1 +c_2\, \lambda)&\mbox{if}\; \xi=1/4\\ \\
e^{-\lambda/2}\left(c_1 \cos\left(\lambda\sqrt{\xi-1/4}\right)+c_2
  \sin\left(\lambda \sqrt{\xi-1/4}\right)\right) &\mbox{if}\;\xi>1/4 
\end{array}
\right.
\ee
where $c_{1,2,3}$ are integration constants determined by the
IBC. Since $\ve(\lambda)$ and 
$\overset{\circ}{\ve}_\bot(\lambda)$ converge to $0$ for large $\lambda$ in all cases, the
critical solution is stable under small perturbations. In other words, 
the solution arising for arbitrary IBC such that $\ve(0)$, $\overset{\circ}{\ve}(0)$ and  $\overset{\circ}{\ve}_\bot(0)$ are small is attracted toward the critical one, $z\equiv z_c$, $\dot\phi_\bot \equiv 0$. The
behavior of $\ve(\lambda)$ is oscillating with damping when $0<\xi<1/4$, and
exponentially convergent when $\xi\ge 1/4$. The special value $\xi=1/4$
corresponds to  $n_V/n_T \simeq -0.0246$.

\subsection{Global attraction to the RDS}
\label{constantz2}

\begin{figure}[h!]
\begin{center}
\vspace{.3cm}
\includegraphics[height=5.5cm]{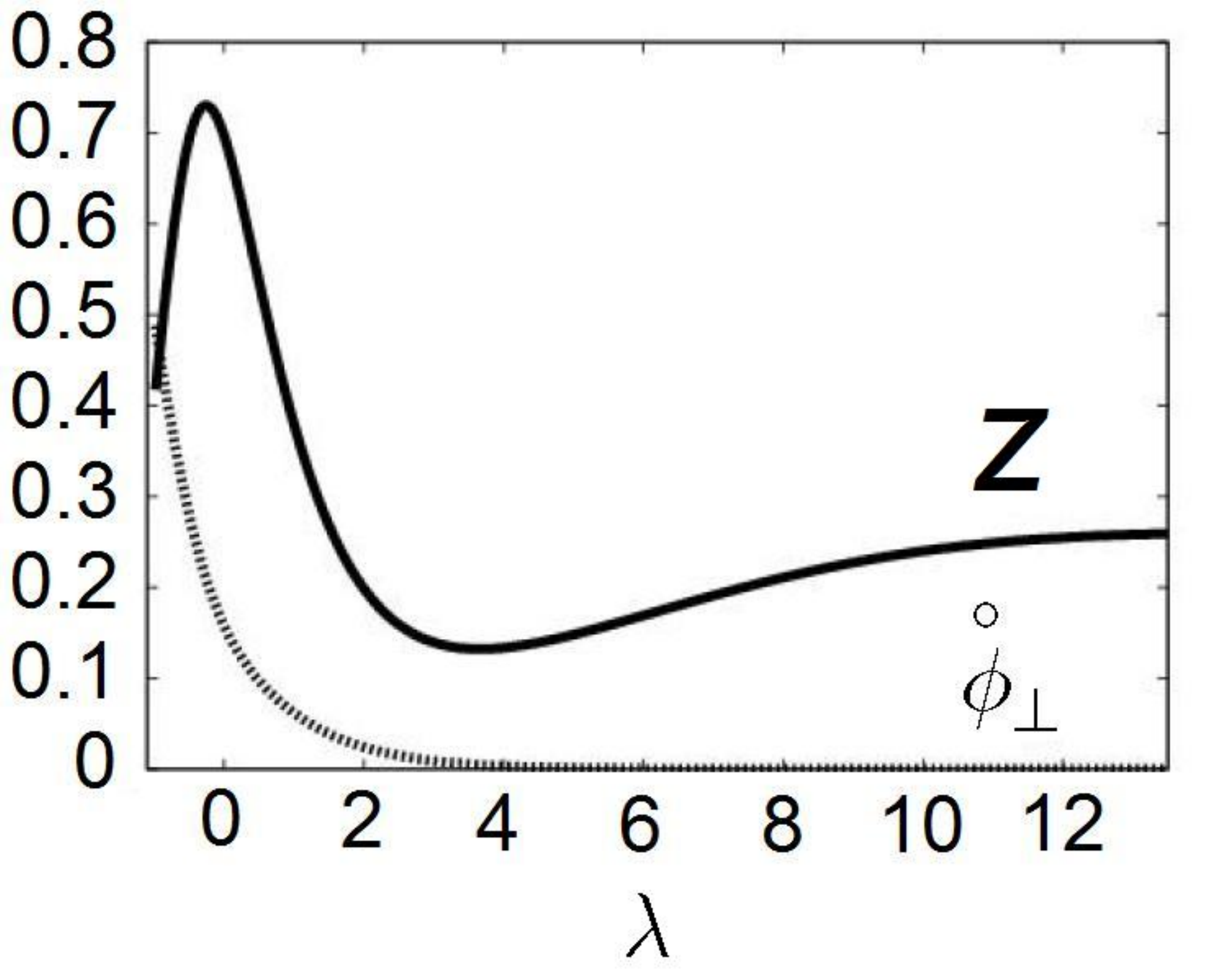}
\vspace{-.3cm}
\caption{\footnotesize \em Examples of damping oscillations of $z(\lambda)$ (solid curve) and convergence to zero of $\overset{\circ}\phi_\bot(\lambda)$ (dotted curve) illustrating the dynamical attraction toward the critical solution $z\equiv z_c\simeq 0.272$, $\dot\phi_\bot\equiv 0$, obtained for $n_V/n_T=-0.02$. The initial conditions are $(z_0, \overset{\circ}{z}_0, \overset{\circ}{\phi}_{\bot 0})=(0.4, 0.8, 0.5)$.}  
\vspace{-.4cm}
\label{fig_numresult}
\end{center}
\end{figure}

Since we have shown the existence of a local basin of attraction around the
critical solution given in equations (\ref{prop}) and (\ref{sol cri}), we want to know if this property is valid for more generic IBC. Actually, any initial values $(z_0,
\overset{\circ}{z}_0, \overset{\circ}{\phi}_{\bot 0})$ at $\lambda = 0$ are
allowed, as long as the positivity of the right hand side of the Friedmann Eq. (\ref{eqh}) is
guaranted\footnote{IBC such that the right hand side of (\ref{eqh}) is negative correspond to solutions in Euclidean time.}, \ie
\be
\label{ic}
{r(z_0)\over 3-\K(z_0, \overset{\circ}{z}_0, \overset{\circ}{\phi}_{\bot 0})}\ge 0\, .
\ee
We have studied numerically the system (\ref{syseq}) for generic IBC satisfying the previous bound. Implementing in the code a positive increment for $\lambda=\ln a$ simulates the phases of the evolutions where the scale factor increases. We observe that the right hand side of (\ref{eqh}) never changes of sign. This means that  the scale factor $a(t)$ is monotonic\footnote{In general, a change of sign in the Friedmann equation during a simulation would be a numerical artifact meaning that $H$ actually vanishes when the scale factor reaches an extremum.}. We always find that the solutions satisfy  $z(\lambda)\to z_c$ and $\overset{\circ}{\phi}_\bot(\lambda)\to 0$ as $\lambda\to +\infty$, after possible oscillations. 
To illustrate the
convergent behavior for large oscillations, we present a numerical
example on Fig. \ref{fig_numresult}. It is obtained for $n_V/n_T=-0.02$ that
corresponds to $z_c\simeq 0.272$, with IC $(z_0, \overset{\circ}{z}_0,
\overset{\circ}{\phi}_{\bot 0})=(0.4, 0.8, 0.5)$. 
We conclude that for generic  IBC, the expanding evolutions are attracted by the RDS that satisfies $z\equiv z_c$, $\dot \phi_\bot \equiv 0$ \ie $3H^2=c_r^2/a^4$ in
Eq. (\ref{prop}). 

The local behavior (\ref{solline}) also shows that the trajectory $z\equiv z_c$, $\dot \phi_\bot \equiv 0$ is repulsive, when $\lambda$ decreases \ie for evolutions where the scale factor is contracting. The Big Crunch solution obtained by time reversal on (\ref{sol cri}) is thus only formal, since it is highly unstable under small fluctuations. What we observe by numerical simulations is that for generic IBC, when we choose a negative increment for $\lambda$, then either $z\to +\infty$, or $z$ oscillates with a greater and greater amplitude. To understand better the dynamics of the contracting evolutions, we are thus led to analyze carefully the system in the regime $z\gg 1$.


\subsection{Attraction to Big Crunch eras}
\label{runaway}

We have seen that when the potential (\ref{r4p}) admits a minimum $z_c$ ({\em Case (b)}), the expanding cosmological evolutions for arbitrary IBC satisfy $z(t)\to z_c$. The shape of the potential in {\em Case (c)} of Fig. \ref{fig_V} suggests that when $n_V\ge 0$, the dynamics could admit  a run away behavior $z(t)\to +\infty$. We also noticed at the end of the previous subsection that the contracting evolutions in {\em Case (b)} involve the large and positive value regime of $z$. For these reasons, we need to study the system for $z\gg 1$.  In this limit the thermal effects ${\rm \cal O} (T¬^4)$ are sub-dominant compare to the supersymmetry breaking effects ${\rm \cal O} (M^4)$.  The expansion of $p(z)$ contains two monomials in $e^z$, 
\be
\label{zbigg}
p(z)=n_V e^{4z} c_5^o+(n_Tc_4^o+n_Vc_4^e)+\cdots\quad \where\quad c_4^e={\Gamma(5/2)\over \pi^{5/2}}\, {2\over
  3}{\sum_m}' {1\over (2m)^4}\, , 
  \ee
and $c_5^o$, $c_4^o$ are defined in Eqs (\ref{V+inf}) and (\ref{V-inf}). The dots stand for exponentially suppressed terms $\O(e^{-e^z})$. It follows that $\A_z(z)\simeq 1$ and the conservation of the energy-momentum Eq. (\ref{phirond}) becomes
\be
\label{aT}
\overset{\circ}{z}=\overset{\circ}{M}/M+1\qquad \Longrightarrow \qquad {\dot T\over T}=-H\qquad \mbox{\ie} \qquad aT=a_0T_0\, ,
\ee
where $a_0T_0$ is a positive constant depending on the IBC. From the expansion of $p(z)$ for large $z$ in Eq. (\ref{zbigg}), it is clear that the behavior of the system is drastically different when  $n_V\neq 0$ and when  $n_V=0$.

\noindent $\bullet$ {\Large \bf {\em n}$_{\! \mbox{\bf \em \small V}}$ $\neq$ 0, {\em z} $\gg$ 1 :}

\noindent We have $\rho\simeq -P\simeq -M^4n_Vc_5^o$. The Friedmann Eq. (\ref{eq1}) and the linear sum of Eqs (\ref{eq1}) and  (\ref{eq2}), together with the matter field 
Eqs (\ref{eq3}) and (\ref{eq4}), become
\begin{eqnarray}
\label{eq1inf} 
&&3H^2={1\over 2} \dot\phi^2+{1\over 2} \dot \phi_\bot^2-M^4n_Vc_5^o\, ,\\ 
\label{eq2inf}
&&\dot H+3H^2 = -M^4n_Vc_5^o\, , \\
\label{eq3inf} 
&&\alpha \ddot{\phi}+3H\alpha\dot\phi=4\alpha^2 M^4n_Vc_5^o\; ,\qquad \ddot{\phi}_\bot+3H\dot\phi_\bot=0\, .
\end{eqnarray}
The above equations  would also arise if we had considered the system at zero temperature. This is due to the fact that when temperature effects are not taken into account, the sources of Einstein gravity always satisfy $\rho=-P$ equal to minus the effective potential (see Eq. (\ref{rho})). However, this does not mean that in our case we shall find that  $T(t)\to 0$. Actually, the regime $z \gg 1$ \ie $M\gg T$ corresponds to  thermal effects screened by radiative corrections, even if the temperature can be large.
Combining  Eq. (\ref{eq2inf}) and Eq. (\ref{eq3inf}) one determines the scalar fields in terms of the scale factor, (with $\alpha=\sqrt{3/2}$),
\be
\label{cphi}
\alpha\dot\phi=-4\alpha^2H+{c_\phi\over a^3}\qquad  \Longrightarrow\qquad   M={e^{c_\phi\int^t{dt'\over a(t')^3}}\over a^6}\qquad \and \qquad \dot\phi_\bot=\sqrt{2}\, {c_\bot\over a^3}\, ,
\ee
where $c_\phi$ and $c_\bot$ are constants depending on the IBC.  The Friedmann 
Eq. (\ref{eq1inf})  takes then the form
\be
\label{Fr}
3H^2={1\over 3}\left(6H-{c_\phi\over a^3}\right)^2+{c_\bot^2\over a^6}-M^4n_Vc_5^o\, .
\ee
We will consider separately the cases $c_\phi = 0$ and $c_\phi \neq 0$.

\noindent {\large \bf {\em (i) c}$_{\bf \phi}$ $=$ 0 :}
\be
M=M_0\left({a_0\over a}\right) ^6 , ~~   3H^2=-{c_m^2\over a^6}+{c_M^2\over a^{24}},\quad {\rm with}\quad c_m={c_\bot^2\over 3}\; , \quad c_M^2={n_V\over 3}\, M_0^4 c_5^o a_0^{24}\, ,
\ee
where $M_0a_0^6$ is a positive constant. This solution exists only for $n_V>0\,$. If we also have  $c_\bot =  0$, up to time reversal, the evolution describes a Big Crunch occurring at a finite time $t_\BC$, 
\be
\label{12}
a(t)= a_0\, (t_\BC-t)^{1/12} \times (16M_0^4n_Vc_5^o)^{1/24}\; , \qquad t< t_\BC\, .
\ee
Since $a\sim 1/T$, it follows that  $e^z\propto aM\propto 1/a^5$, which is large if $t\lsim t_\BC$ and implies a run away of $z$, consistently with our hypothesis $z\gg 1$. However, close to the Big Crunch, the scale $M$ (as well as $T$, even if $T\ll M$) is formally diverging.  This implies that the present analysis has to be restricted to the regime where $M< M_H$ \ie $z<(5/6)\ln M_H$, where $M_H$ is the supersymmetry breaking scale above which a Hagedorn-like transition occurs. To bypass this limit, one could try to extend our study  to Hagedorn singularity free models \cite{GravFluxes}. 

 For $c_\bot\neq 0$, the time can be expressed as a function of the scale factor,
\be
\label{sol1}
t(a)=\pm \, t_\BC\int_{a/a_\max}^1{x^{11}dx\over \sqrt{1-x^{18}}}\quad \where \quad  t_\BC={\sqrt{3}\over \abs c_M\abs }\, a_\max^{12} \; ,  \quad \forall a \leq a_\max=\left\abs {c_M\over c_m}\right\abs^{1/9}\, .
\ee
Formally, this solution describes a Big Bang  at $t=-t_\BC$ that initiates an expansion era. The latter stops when the scale factor reaches its maximum $a_\max$ at $t=0$. The universe then  contracts till $t=t_\BC$, where a Big Crunch occurs. As before, the last era of this evolution can be trusted for $t\lsim t_\BC$, where the solution (\ref{sol1}) is asymptotic to (\ref{12}). Thus, we have shown that for IBC such that $z_0\gg 1$ and $c_\phi=0$, the dynamics is  attracted to the MDS described by the  Big Crunch solution of Eq. (\ref{12}) with $c_\phi=c_\bot=0$  and $3H^2=c_M/a^{24}$. 

\noindent {\large \bf {\em (ii) c}$_{\bf \phi}$ $\neq$ 0 :}
 
\noindent We consider  IBC such that $z_0\gg 1$, with generic  $c_\phi \neq 0$. The sign of $n_V\neq 0$ is arbitrary and we want again to find the basins of attraction of the dynamics. 
Instead of Eq. (\ref{eq1inf}), it is more convenient to use the combination of Eq. (\ref{eq1inf}) and  Eq. (\ref{eq2inf}),
\be
\label{deriFried}
\dot H = -{1\over 2} \dot\phi^2-{1\over 2} \dot \phi_\bot^2\, ,
\ee
that becomes, utilizing the  expressions (\ref{cphi}) for  $\dot \phi$ and $\dot \phi_\bot$:
\be
\label{Fr2}
\dot H=-{1\over 3}\left(6H-{c_\phi\over a^3}\right)^2-{c_\bot^2\over a^6}\, .
\ee
Introducing the quantity  $\sigma$,
\be 
\label{sigma}
\sigma={9\over 4c_\phi}\, {a^2\dot a }~={~3\over 4c_\phi}\, (a^3)^{\mbox{$\cdot$}}~=~{9\over 4c_\phi} {a^3 H}\, ,
\ee
Eq. (\ref{Fr2}) takes the form:
\be
\label{eq}
{\sigma \, d\sigma\over \sigma^2-\sigma+{3\over 16}(1+3(c_\bot/c_\phi)^2)}=-9 \, {da\over a}\, ,
\ee
 valid in the $z\gg 1$ limit. In the same limit the field equations  become: 
\be
\label{sys+inf}
\left\{
\begin{array}{l}
\displaystyle
{1\over {1\over 3}(\overset{\circ}{M}/M)^2+{1\over 2}\, \overset{\circ}{\phi}_\bot\!\!\!{}^2-3}(\overset{\circ}{M}/M)^\circ -\overset{\circ}{M}/M-6=0\\ 
\displaystyle {1\over {1\over 3}(\overset{\circ}{M}/M)^2+{1\over 2}\, \overset{\circ}{\phi}_\bot\!\!\!{}^2-3}\overset{\circ\circ}{\phi}_\bot -\overset{\circ}{\phi}_\bot=0
\end{array}
\right.
\ee
where the signs of $n_V$ and the fractions in front of the second derivatives must be the same, so that the right hand side of Eq. (\ref{eqh}) is positive. This condition is required to study the evolutions in real time (and not Euclidean). This differential system is easily shown to be equivalent to
\be
\label{eqq}
\left\{
\begin{array}{l}
\displaystyle
3(\overset{\circ}{M}/M)^\circ=(\overset{\circ}{M}/ M+6) \P(\overset{\circ}{M}/M) \\ 
\displaystyle\overset{\circ}\phi_\bot=\sqrt{2}\, {c_\bot\over c_\phi}\, ( \overset{\circ}{M}/M+6)
\end{array}
\right.
\ee
where $c_\bot/c_\phi$ is an arbitrary constant denoted this way to make contact with the conventions of 
Eq. (\ref{cphi}), and the polynomial
\be
\label{P}
 \P(x)=[1+3(c_\bot/ c_\phi)^2]\, x^2+36\, (c_\bot/ c_\phi)^2\, x+9\, [12(c_\bot/ c_\phi)^2-1]\mbox{ is of the sign of $n_V$.}
\ee
Using the first equation in (\ref{cphi}), one obtains 
\be
\sigma={9\over 4}\, {1\over \overset{\circ}{M}/ M+6}\, ,
\ee
that can be used to identify consistently (\ref{eq}) with the first equation of  (\ref{sys+inf}). By the way, the real time condition translates to the condition that $n_V$ and the denominator of the left hand side of (\ref{eq}) have a common sign.  We are going to solve this equation separately for $n_V>0$ and $n_V<0$.  

{\em \large - For $n_V>0$ : }

\begin{figure}[h!]
\begin{center}
\vspace{.3cm}
\includegraphics[height=5cm]{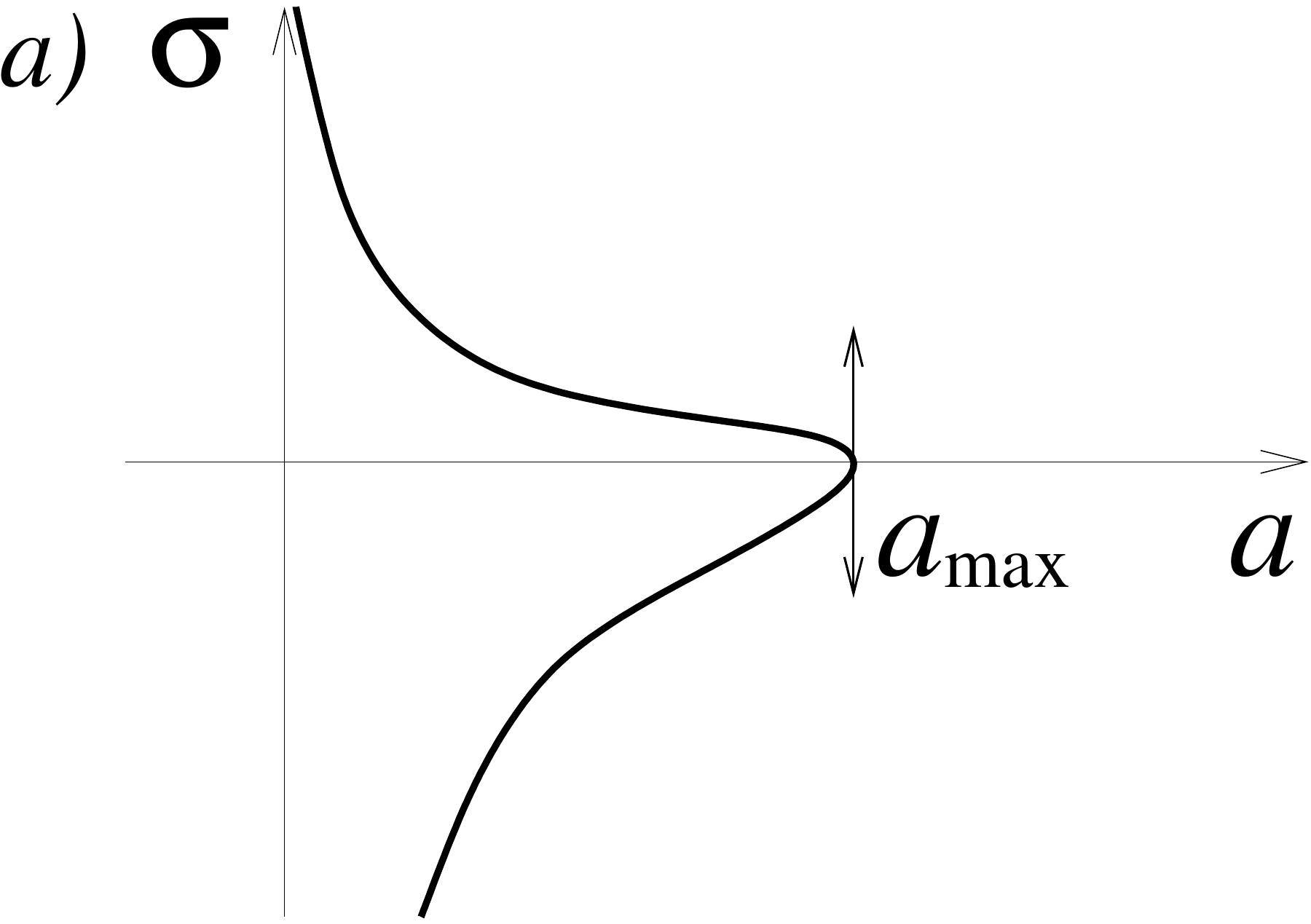}
\qquad \qquad
\includegraphics[height=5cm]{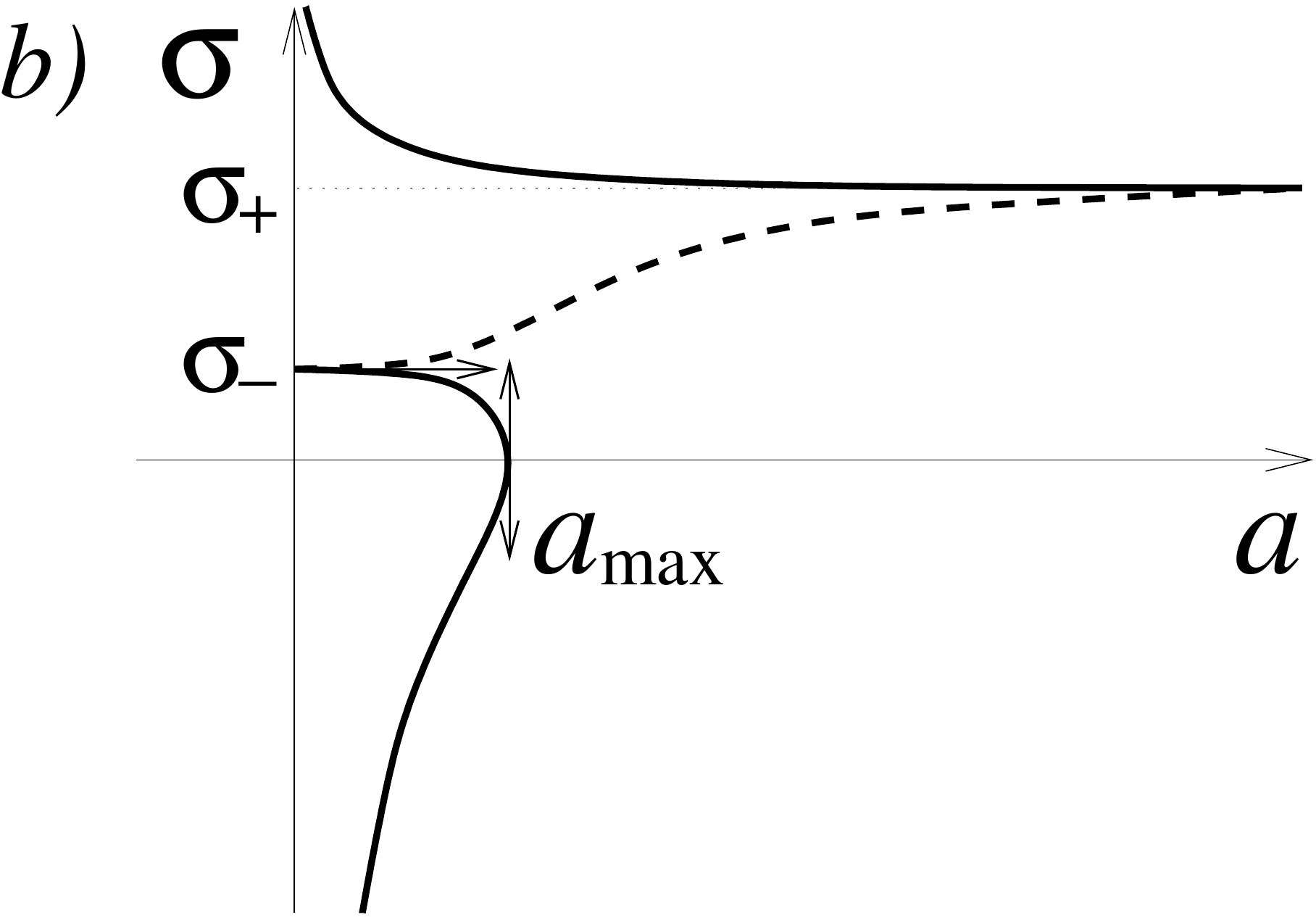}
\caption{\footnotesize \em $\sigma$, defined in Eq. (\ref{sigma}) versus the scale factor $a$. Fig. a) corresponds to $(c_\bot/c_\phi)^2>1/9$ and Fig. b) corresponds to $(c_\bot/c_\phi)^2\le 1/9$. The solid lines are associated to the case $n_V>0$, while the dashed one is associated to the case $n_V<0$. }  
\vspace{-.4cm}
\label{sigma_vs_a}
\end{center}
\end{figure}

{\em i)} When $(c_\bot/c_\phi)^2>1/9$, the denominator of (\ref{eq}) has no real root.  Integrating, one obtains
\be
\label{soli}
{a\over a_0}={\exp\left({-{2\over 9\sqrt{9(c_\bot/c_\phi)^2-1}}\tan^{-1}\left({4\sigma-2\over \sqrt{9(c_\bot/c_\phi)^2-1}}\right)}\right)
\over \left(  \sigma^2-\sigma+{3\over 16}(1+3(c_\bot/c_\phi)^2) \right)^{1/18}}\, ,
\ee

\noindent
that is used to draw $\sigma$ versus $a$ on Fig. \ref{sigma_vs_a}{\it a}. 
Suppose $c_\phi>0$. If at some time $\sigma >0$, the definition of $\sigma$ in 
Eq. (\ref{sigma}) implies that $a$ increases. Thus, the scale factor reaches a maximum $a_\max$ where $\sigma$ vanishes. We then pass into the $\sigma <0$ lower half plane where $a$ decreases and converges to 0. Since  $\sigma \to -\infty$ and $a\to 0$, one has $\dot a\to -\infty$, which is interpreted as a Big Crunch occurring at a finite time $t_\BC$. 
Similar arguments apply when $c_\phi<0$ : If $\sigma <0$, the scale factor increases up to $a_\max$ before converging to 0. This corresponds to a Big Crunch occurring at some finite time $t_\BC$. 
To summarize, when the IBC are such that $c_\phi >0$ ($c_\phi <0$), the system follows the curve on Fig. \ref{sigma_vs_a}{\it a} from up to down (down to up). The universe starts from a Big Bang, reaches a maximum size, and ends with  a Big Crunch. Of course, only the parts of this evolution consistent with the hypothesis $z\gg 1$ can be trusted. In particular, when we approach $t_\BC$, one has $\sigma\to -s\infty$, where $s=\mbox{sign}(c_\phi)$,  and Eq. (\ref{soli}) implies,
\be
\label{12b}
a(t)\simeq (t_\BC-t)^{1/12} \times ((16/3)\abs c_\phi\abs \,a_0^9 \,e^{s9\pi/2})^{1/12}\; , \qquad t\lsim t_\BC\, .
\ee
This result can be used to show that the integral in $M$ in Eqs (\ref{cphi}) is bounded when $t\to t_\BC$ so that a consistent run away $e^z\propto 1/a^5\to +\infty$ is found. An important remark then follows. The behavior (\ref{12b}) implies that $c_\phi/a^3$ is dominated by $H$ in Eq. (\ref{cphi}), showing that the dynamics for $(c_\bot/c_\phi)^2>1/9$ is always attracted by the solution for $c_\phi=c_\bot=0$.

{\em ii)} When $(c_\bot/c_\phi)^2\le 1/9$, the left hand side of Eq. (\ref{eq}) has two real roots. Integrating, one has
\begin{eqnarray}
\mbox{for }\left({c_\bot\over c_\phi}\right)^2< {1\over 9}&: & \quad {a\over a_0}={\abs\sigma-\sigma_-\abs^{\sigma_-\over 9(\sigma_+-\sigma_-)}\over \abs\sigma-\sigma_+\abs^{\sigma_+\over 9(\sigma_+-\sigma_-)}}\; ,   \quad \sigma_\pm={1\over 2}±\pm {1\over 4}\sqrt{1-9(c_\bot/c_\phi)^2}\, ,\\
\mbox{for }\left({c_\bot\over c_\phi}\right)^2= {1\over 9}&: & \quad {a\over a_0}={e^{1\over 18(\sigma-1/2)}\over \abs \sigma -{1\over 2}\abs^{1/9}}\; ,   \quad \sigma_\pm={1\over 2}\, ,
\end{eqnarray}
where $\sigma_-<\sigma <\sigma_+$. Fig. \ref{sigma_vs_a}{\it b} shows $\sigma$ as a function of $a$.\footnote{To be precise, the slope at the point $(a,\sigma)=(0,\sigma_-)$ is infinite when $(1-4/361)/9<(c_\bot/c_\phi)^2\le 1/9$, finite when  $(c_\bot/c_\phi)^2=(1-4/361)/9$ and vanishing when $(c_\bot/c_\phi)^2<(1-4/361)/9$.} When $c_\phi>0$, the arguments used in case {\em i)} apply on the branch $\sigma<\sigma_-$, where the system evolves from up to down. The scale factor grows up to $a_\max$ before converging to 0. Similarly, when $c_\phi<0$ and the system follows the branch  $\sigma>\sigma_+$, the scale factor converges to 0. In both cases, a Big Crunch occurs at $t_\BC$ and the behavior (\ref{12b}), with $s=0$, is valid. The dynamics is again attracted by the solution associated to $c_\phi=c_\bot=0$. 

Some new features occur for $c_\phi <0$, when $\sigma<\sigma_-$, (see Fig.  \ref{sigma_vs_a}{\it b}). The scale factor increases up to its maximum $a_\max$ before converging to 0, while $\sigma\to \sigma_-$, implying $\dot a\to -\infty$. A Big Crunch at a finite time $t_\BC$ occurs but since $\sigma\to \sigma_-$, one has
\be
\label{3}
a(t)\simeq (t_\BC-t)^{1/3} \times ((4/3)\abs c_\phi\abs \sigma_-)^{1/3}\; , \qquad t\lsim t_\BC\, .
\ee
However, this behavior can be trusted as long as $e^z\propto aM\propto  \abs t_\BC-t\abs^{{3\over 4\sigma_-} -{5\over 3}}\to +\infty$. This condition implies that the scaling $(t_\BC-t)^{1/3}$ exists when  
\be
\label{cond3}
c_\phi<0\qquad \and \qquad {1\over 9}\, (1-1/25)<\left({c_\bot\over c_\phi}\right)^2\le{1\over 9}\, .
\ee
In this case, the evolution is not attracted by the solution $c_\phi=c_\bot=0$. 
Actually, Eq. (\ref{cphi}) can be rewritten as 
\be
\alpha \phi\simeq -\left(2-{3\over 4\sigma_-}\right)\, \ln\abs t_\BC-t\abs+c\to +\infty\qquad \when \qquad t\to t_\BC\, ,
\ee
so that the integration constant $c$ is negligible. There is thus a new basin of attraction, to the solution (\ref{3}) with $c=0$ \ie such that  
\be
\label{attrac1/3}
3H^2\simeq {c_m^2\over a^6}\qquad \where\qquad c_m={1\over 27}\, c_\phi^2\, \sigma_-^2\; , \quad c=0\, .
\ee 
As for the previous Big Crunch behavior, the divergence of $M$ (and $T$) is only formal since the present analysis supposes $M<M_H$, where a Hagedorn-like transition occurs. On the contrary, for $(c_\bot/c_\phi)^2$ smaller than the lower bound of the range (\ref{cond3}),  we formally have $e^z\propto  \abs t_\BC-t\abs^{{3\over 4\sigma_-} -{5\over 3}}\to 0$, which shows that the system actually exits from the regime $z\gg 1$ we started with. Similarly, if  $c_\phi >0$ and the system evolves along  the branch $\sigma>\sigma_+$ of Fig. \ref{sigma_vs_a}{\it b}, one has  $a\to +\infty$ and $\sigma\to \sigma_+$. This implies $a\simeq t^{1/3} \times ((4/3)c_\phi\sigma_+)^{1/3}$ and $t\to +\infty$. However, for large positive $z$, this implies $e^z\propto aM\propto t^{{3\over 4\sigma_+} -{5\over 3}}\to 0$, which shows that the system actually exits in the regime $z\gg1$. 

{\it \large Summary for $n_V>0$ :} 

\noindent We have shown analytically that when $z_0\gg 1$, either $z\to +\infty$ or $z$ quits the regime $z\gg 1$. To study the dynamics at finite $z$, we use again a code that implements  Eqs (\ref{syseq}). The phases of expansion are simulated when a positive increment for $\lambda$ is chosen. For generic IBC, we observe that the right hand side of the Friedmann equation (\ref{eqh}) changes sign at some finite $a=a_\max$. This actually means that the scale factor has reached a maximum, (the Hubble parameter $H$ vanishes), and that  the evolution enters into a phase of contraction. Such eras are simulated by choosing a negative increment for $\lambda$ and we observe that for  generic  IBC, $z$ ends by running away. 
As a conclusion, the generic IBC admit two basins of attraction to MDS. The evolutions converge either to the solution where $c_\phi=c_\bot=0$ \ie such that $3H^2=c_M^2/a^{24}$, or to the solution with $c=0$ and $3H^2=c_m^2/a^{6}$.

{\em \large - For $n_V<0$ : }

\noindent
For IBC such that $z_0\gg 1$, the real time condition translates to $\sigma_-<\sigma<\sigma_+$, (see Fig. \ref{sigma_vs_a}{\it b}).  If $c_\phi>0$, the definition of $\sigma$ in Eq. (\ref{eq}) reaches $\dot a>0$. It follows that $a\to +\infty$ and $\sigma\to \sigma_+$. Thus, $a\simeq t^{1/3} \times ((4/3)c_\phi\sigma_+)^{1/3}$ where $t\to +\infty$, and $e^z\propto aM\propto t^{{3\over 4\sigma_+} -{5\over 3}}\to 0$. We conclude that the system exits the regime $z\gg1$. When $-1/15<n_V/n_T<0$ ({\em Case (b)}), this result is compatible with Sect. \ref{constantz}, where we found that the expanding evolutions satisfy $z\to z_c$.

If on the contrary $c_\phi<0$, one has $\dot a<0$. It follows from Fig. \ref{sigma_vs_a}{\it b} that $a\to 0$ and  $\sigma\to \sigma_-$, so that $\dot a\to -\infty$. The dynamics is attracted by the Big Crunch solution (\ref{3}) \textit{i.e.} (\ref{attrac1/3}) that can be trusted when the conditions (\ref{cond3}) are satisfied. For smaller values of $(c_\bot/c_\phi)^2$, the system exits the regime $z\gg 1$. These conclusions are compatible with the numerical simulations of the contracting evolutions described at the end of Sect. \ref{constantz2}, for generic IBC  in {\em Case (b)}. We found there that either $z\to +\infty$, as expected from the existence of the Big Crunch solution (\ref{3}), or $z$ oscillates with a larger and larger amplitude, implying that when $z$ is large and positive, it can exit the regime $z\gg 1$.

\noindent $\bullet$ {\Large \bf {\em n}$_{\! \mbox{\bf \em \small V}}$ $=$ 0, {\em z} $\gg$ 1 :}

\noindent We argued in Section \ref{constantz} that when $-1/15<n_V/n_T<0$, the expanding evolutions are such that $z$ is attracted by a critical value $z_c$. Since $z_c  \to +\infty$ when $n_V/n_T \to 0_-$, we expect the expanding solutions for $n_V=0$ to satisfy $z(t)\to +\infty$. Actually, this run away may be natural since the potential in Eq. (\ref{r4p}) decreases linearly for large positive $z$, $V\sim -z n_T c_4^o$.  
In fact, a numerical study of the differential system (\ref{syseq}) with $n_V=0$ confirms that for generic IBC, $z$ enters the regime $z\gg 1$ when we simulate the expanding solutions. On the other hand, for the evolutions where the scale factor decreases, we find either $z\to +\infty$ or $z$ oscillates with a larger and larger amplitude. In all these cases, we are thus invited to consider analytically the system in the regime $z\gg1$.

In this limit, the thermal effective potential for $\phi$ in Eq. (\ref{eq3}) vanishes (up to exponentially small terms in $z$). Both $\phi$ and $\phi_\bot$ are flat directions, and their kinetic energies involve two constants $c_\phi$, $c_\bot$, determined by the IBC,   
\be
\label{kine}
\alpha\, \dot\phi={c_\phi\over a^3}\; , \quad \dot\phi_\bot= \sqrt{2}\, {c_\bot\over a^3}\ .
\ee
The energy density in the Friedmann Eq. (\ref{eq1}) is $\rho\simeq 3T^4 n_Tc_4^o$ and,  using (\ref{aT}), one has 
\be
\label{46}
3H^2= {c_r^2\over a^4}+{c_m^2\over a^6}\qquad \where\qquad c_r^2= 3n_Tc_4^o(a_0T_0)^4\; , \quad c_m^2=c_\bot^2+{c_\phi^2\over 3}\, .
\ee
This equation admits two monotonic solutions (see Eq. (\ref{expan})), mapped to one another under time reversal. However, one can only trust them as long as  $z\gg 1$.

- For  the expanding evolutions, since $a(t)\to +\infty$ when $t\to +\infty$, Eq. (\ref{kine}) implies that $M(t)$ converges to a constant $M_0$. It follows that $e^z=M/T\propto a \to +\infty$, showing the validity of the solution for large enough time. We conclude that for generic IBC, the expanding evolutions are attracted to the RDS, Eq. (\ref{sol cri}),  with $M=M_0$.

- For the contracting evolutions, depending on $c_\phi$ and $c_\bot$, we have 
\be
\label{6BC}
a(t)\simeq (t_\BC-t)^{1/3}\times (3c_m^2)^{1/6}\; , \qquad M\simeq M_0\left(1-{t\over t_\BC}\right)^{-{1\over  [1+3(c_\bot/c_\phi)^2]^{1/2}}}  \; , \qquad t\lsim t_\BC
\, ,
\ee
if the constraints 
\be
c_\phi>0\qquad \and \qquad \left({c_\bot\over c_\phi}\right)^2<{8\over 3}
\ee
are satisfied, (so that $z\to +\infty$ when $t\to t_\BC$). Alternatively, if $c_\phi=c_\bot=0$, the solution
\be
\label{4BC}
a(t)\simeq \sqrt{t_\BC-t}\times \left({4\over 3}\, c_r^2\right)^{1/4}\; , \qquad M\simeq M_0 
\ee
is also consistent with a run away of $z$. For all other values of $c_\phi$ and $c_\bot$, $z$ exits the regime $z\gg 1$. Our conclusions for the contracting evolutions are that  there are two sets of generic IBC. The first  gives rise to a run away of $z$ that corresponds to  the Big Crunch solutions (\ref{6BC}) or (\ref{4BC}). The second yields an oscillation regime in $z$, with larger and larger amplitude. 


\section{Susy breaking involving {\em n} = 2 internal directions}

In this section we extend our analysis to models where the supersymmetry  breaking is induced  by 
$n=2 $  Scherk-Schwarz circles $S^1(R_4)\times S^1(R_5)$ in the 4$^{\mbox{\footnotesize th}}$ and 5$^{\mbox{\footnotesize th}}$ internal 
dimensions as in Ref. \cite{Cosmo-2}. 
Thus, we are  considering  the heterotic or type II superstrings on
\be
S^1(R_0)\times T^3\times S^1(R_4)\times S^1(R_5)\times {\cal M}_4\, , 
\ee
where ${\cal M}_4$ is either $T^4$ or $K3\sim T^4/\Z_2$.  As shown in \cite{Cosmo-2}, such models allow cosmological evolutions that describe radiation dominated eras, with stabilized complex structure modulus $R_5/R_4$\cite{Cosmo-2}. The present analysis could certainly be extended to the other classes of models described in  \cite{Cosmo-2}, where asymmetric Scherk-Schwarz compactifications in the directions 4 and/or 5 are considered. These models would involve run away solutions in the spirit of the {\em Cases (c)} and {\em (a)} for $n=1$.  

$R_0$, $R_4$, $R_5$ are restricted to be  large enough in order to avoid Hagedorn-like phase transitions. In this regime,  the free energy and  pressure are computed in Ref. \cite{Cosmo-2} :
\be
\label{P2mod}
P=T^4\, p(z,Z)\quad \where \quad e^z={M\over T}\; , ~~T={1\over 2\pi R_0\sqrt{\Re S}}\; , ~~M={1\over 2\pi \sqrt{R_4R_5}\sqrt{\Re S}}\; , ~~ e^Z={R_5\over R_4}\, .
\ee
$T$ is the temperature in the Einstein frame,  $M$ is the supersymmetry breaking scale which is given in terms of  the geometric mean of the moduli $R_4$ and $R_5$, and $Z$ parametrizes the complex structure modulus $R_5/R_4$. The quantity $p(z,Z)$ is a linear sum of functions, with model-dependent integer coefficients,
\be
\label{pn}
p(z,Z)=n_{T}\, p_{100}(z,Z)+n_{010\, }p_{010}(z,Z)+n_{001}\, p_{001}(z,Z)+n_{111}\, p_{111}(z,Z)\, ,
\ee
where $n_{T}$ is again the number of massless states and 
\be
\label{fonc3}
p_{\tg_0\tg_4\tg_5}={2\over \pi^3}\sum_{\tm_0, \tm_4,\tm_5}{e^{4z}\over \left[(2\tm_0+\tg_0)^2e^{2z}+(2\tm_4+\tg_4)^2e^{-Z}+(2\tm_5+\tg_5)^2e^{Z}\right]^3}\, .
\ee

\subsection{Gravitational and moduli equations}
The low energy effective action (\ref{action}) with the kinetic term of $R_5$ taken into account can be written in terms of redefined fields as :
 \be
S = \int d^4x \sqrt{-g}\left[ {1\over 2}\, R-{1\over 2}\left((\partial
    \phi)^2+(\partial \phi_\bot )^2+{1\over 2}(\partial Z)^2\right)+P \right]\, ,  
\ee
where
\be
\phi:= \phi_D-\ln \sqrt{R_4R_5}\; , \quad \phi_\bot := \phi_D+\ln \sqrt{R_4R_5}\, ,\quad P= M^4\, e^{-4z}p(z,Z)\: , \quad M={e^{\phi}\over 2\pi}\, . 
\ee
For evolutions satisfying the metric ansatz (\ref{FRW}), the Friedmann equation is
\be
\label{eq1bis}
3H^2={1\over 2} \dot\phi^2+{1\over 4} \dot Z^2+{1\over 2} \dot \phi_\bot^2+T^4\, r \quad \where \quad r(z,Z)=3p-p_z\, ,
\ee
while the equation for the scale factor,
\be
\label{eq2bis}
2\dot H+3H^2=-{1\over 2} \dot \phi^2-{1\over 4} \dot Z^2-{1\over 2} \dot \phi_\bot^2-T^4p\, ,
\ee
can be replaced by the conservation of the energy-momentum in a form similar to Eq. (\ref{cEbis}),
\be
\label{cEbisbis}
\dot\rho+3H(\rho+P)+\dot\phi \, T^4\, (3p-r)+\dot Z\, T^4\, p_Z=0\, .
\ee
For the scalar fields, one has
\begin{eqnarray}
&&\label{eq3bis}
\ddot{\phi}+3H\dot\phi=T^4\, (3p-r)\, ,\\
&&\label{eq4bis}
\ddot{Z}+3H\dot Z=T^4\, 2p_Z\, ,\\
&&\label{eq5bis}
\ddot\phi_\bot+3H\dot\phi_\bot=0\qquad \Longrightarrow\qquad \dot
\phi_\bot=\sqrt{2}\; {c_\bot\over a^3}\, .
\end{eqnarray}

We proceed by introducing the variable $\lambda$ of (\ref{new t}) as in Section \ref{classi} to express
\be
\label{phirondbis}
\overset{\circ}{\phi}  = \A(z,Z)\overset{\circ}{z}+ \B(z,Z)\overset{\circ}{Z}-1\quad
\where \quad \A(z,Z)={4r-r_z\over 3(r+p)}\; , ~~ \B(z,Z)=-{r_Z+p_Z\over 3(r+p)}\; ,
\ee
that reaches, by integration, 
\be
\label{Ma2}
M={e^{\F(\ln a)}\over a}\qquad\where \qquad \overset{\circ}\F(\lambda):=\A(z,Z)\overset{\circ}z+\B(z,Z)\overset{\circ}Z\, .
\ee
This result can be used to rewrite the Friedmann Eq. (\ref{eq1bis}) in either of the following  forms, 
\be
\label{Friedrondbis}
3H^2={3\over 5}\, H^2\left( (\A\overset{\circ}z+\B\overset{\circ}Z-1)^2+{1\over2}\,  \overset{\circ}Z -1\right)^2+{6\over 5}\, {e^{4\left[\F(\ln a)-z\right]}\over a^4}\, r(z,Z)+{6\over 5}\, {c_\bot^2\over a^6}\, ,
\ee
or
\be
\label{eqhbis}
H^2=T^4\, {r(z,Z)\over 3-\K(z,\overset{\circ}{z},Z,\overset{\circ}{Z},\overset{\circ}{\phi}_\bot)}\quad \where
\quad \K={1\over 2} \left(\A(z,Z)\overset{\circ}{z}+\B(z,Z)\overset{\circ}{Z}-1\right)^2+{1\over 4}\, \overset{\circ}{Z}{}^2+{1\over 2}\, \overset{\circ}{\phi}_\bot\!\!\!{}^2\, ,
\ee  
while the equations for the scalars become
\be
\label{syseqbis}
\left\{
\begin{array}{l}
\displaystyle
{r\over 3-\K}\left(\A\, \overset{\circ\circ}{z}+\B\, \overset{\circ\circ}{Z}+\A_{z}\overset{\circ}z{}^2+(\A_{Z}+\B_{z})\overset{\circ}z\overset{\circ}Z+\B_{Z}\overset{\circ}Z{}^2\right)+{r-p\over 2} (\A\, \overset{\circ}{z}+\B\, \overset{\circ}{Z})+V_z^{(z)}=0\\ \\
\displaystyle {r\over 3-\K}\,
\overset{\circ\circ}Z +{r-p\over 2}\,
\overset{\circ}Z+V_Z^{(Z)} =0 \\ \\
\displaystyle {r\over 3-\K}\,
\overset{\circ\circ}{\phi}_\bot +{r-p\over 2}\,
\overset{\circ}{\phi}_\bot =0 
\end{array}
\right.
\ee
where we have defined
\be
\label{r5p}
V^{(z)}_z(z,Z)={r-5p\over 2}\quad\and\quad V^{(Z)}_Z(z,Z)=-2\, p_Z\, .
\ee

\subsection{Attraction to the RDS}

A wide range of behaviors can emerge from this system of differential equations. However, in the spirit of Section \ref{constantz}, we choose to look for cosmological solutions with stabilized complex structures $e^z=R_0/\sqrt{R_4R_5}$ and $e^Z=R_5/R_4$ \ie satisfying $(z,Z)\equiv (z_c,Z_c)$. From the system (\ref{syseqbis}), such solutions exist if the following conditions are simultaneously  satisfied,
\begin{eqnarray}
\label{C=0}&V_z^{(z)}(z_c,Z_c)=0& \quad \mbox{\ie}\quad 2p(z_c,Z_c)+p_z(z_c,Z_c)=0\, ,\\
\label{D=0}&V_Z^{(Z)}(z_c,Z_c)=0&\quad \mbox{\ie}\quad p_Z(z_c,Z_c)=0\, .
\end{eqnarray}
These equations define two sets of curves in the $(z,Z)$-plane, whose intersections determine the critical points $(z_c,Z_c)$. Supposing such a solution exists, the corresponding cosmological evolution is characterized by $\overset{\circ}\F=0$ in Eq. (\ref{Ma2}) and thus
\be
\label{propbis}
M(t)=e^{z_c}\, T(t)={e^{\F}\over a(t)}\qquad \where \qquad \F=\mbox{cst.}\; ,  \qquad R_5(t)=e^{Z_c}\, R_4(t)\, ,
\ee
together with the Friedmann Eq. (\ref{Friedrondbis}),
\be
\label{cosmoTbis}
3H^2={c^2_r\over a^4}+{c^2_m\over a^6}\qquad \where \qquad
c^2_r=6\, e^{4(\F-z_c)}\, p(z_c,Z_c) \; , \quad
c^2_m={6\over 5}\, c^2_\bot\, .
\ee
The latter is qualitatively identical to the one found for $n=1$ in {\em Case (b)}. The evolution with expanding scale factor is asymptotic to the critical solution $(z,Z)\equiv (z_c,Z_c)$, $c_m=0$, where $a(t)$ is given in (\ref{sol cri}). 

To study the local stability around a critical solution with $c_m=0$, we define small perturbations around it,
\be
z(\lambda)=z_c+\varepsilon(\lambda)\; , \qquad Z(\lambda)=Z_c+E(\lambda)\; , \qquad\overset{\circ}\phi_\bot(\lambda)=\overset{\circ}\varepsilon_\bot (\lambda)\, ,
\ee
and consider the system (\ref{syseqbis}) at first order in $\varepsilon$, $E$ and $\overset{\circ}\varepsilon_\bot$. The latter can be brought into the form
\be
\label{linebis}
\left\{
\begin{array}{llll}
\displaystyle \overset{\circ\circ}{\ve}+\overset{\circ}{\ve}+\xi\,
\ve+\omega\, E=0&\where&\xi=\left. {9p\,(4p-p_{zz})+2p^2_{zZ}\over 2p\,(26p+p_{zz})} \right\abs_{(z_c,Z_c)}\; ,& \omega=\left.{p_{zZ}\, (2p_{ZZ}-9p)\over 2p\, (26p+p_{zz})}\right\abs_{(z_c,Z_c)}\, , \\ 
    \displaystyle \overset{\circ\circ}E+\overset{\circ}E+\Xi\,
E+\Omega\, \ve=0&\where&\Xi=-\left.{p_{ZZ}\over p}\right\abs_{(z_c,Z_c)}\; , & \Omega=-\left.{p_{zZ}\over p}\right\abs_{(z_c,Z_c)}\, ,\\ 
\displaystyle \overset{\circ\circ}{\ve}_\bot+\overset{\circ}{\ve}_\bot=0\, .\end{array}
\right.
\ee
To proceed, we specialize on the set of models whose R-symmetry charges coupled to the winding and momentum numbers in the directions 4 and 5 are identical. For this set, it is shown in \cite{Cosmo-2} that the function $p$ in Eq. (\ref{pn}) takes the form,
\be
\label{psym}
p(z,Z)=n_{T} [p_{100}+p_{111}]+n_V[p_{010}+p_{001}]\; , \quad -1\le {n_V\over n_T}\le 1\, ,
\ee
and that the range of $n_V/n_T$ can be divided in four phases whose frontiers are given by $\{ -0.215,\, -1/31,\, 0\} $\cite{Cosmo-2}.
\begin{itemize}
\item For $n_V/n_T\lsim -0.215$, the constraint (\ref{C=0}) has no solution.
\item For $-0.215\lsim n_V/n_T<-1/31$, the constraint (\ref{C=0}) defines a non trivial curve, while the condition (\ref{D=0}) is satisfied on  the axis $Z=0$. The two loci intersect at a point $(z_c,0)$.  
\item For $-1/31<n_V/n_T<0$, the constraint (\ref{C=0}) defines a non trivial curve, while the condition (\ref{D=0}) corresponds to  3 curves, one of which being the axis $Z=0$. There is a unique intersection point $(z_c,0)$. 
\item For $0<n_V/n_T$, the constraint (\ref{C=0}) has no solution.
\end{itemize}
Thus, a solution to the equations of motion with constant  complex structures $(z_c,Z_c)$ exists when $-0.215\lsim n_V/n_T<0$. 

The class of models we consider have a symmetry $R_4\leftrightarrow R_5$ that translates to $Z\leftrightarrow -Z$. This implies $p_{zZ}(z_c,0)=0$ and 
\be
\xi={9\over 2} \left. {4p-p_{zz}\over 26p+p_{zz}} \right\abs_{(z_c,Z_c)}\; , ~~\omega=0\; , ~~\Xi=-\left.{p_{ZZ}\over p}\right\abs_{(z_c,Z_c)}\; , ~~\Omega=0\, .
\ee
As in the $n=1$ case, one finds numerically that $\xi$ and $\Xi$ are always positive. They increase from 0 to $+\infty$ when $n_T/n_V$ varies from $\simeq -0.215$ to 0, which shows that $\ve(\lambda)$, $E(\lambda)$ (and $\overset{\circ}\ve_\bot(\lambda)$) are always converging to zero. This means that in the class of models under investigation, when a solution $(z_c,Z_c)$ exists, it is  stable under small perturbations. 

A numerical study of the exact system  (\ref{syseqbis}) shows that for generic IBC $(z_0,\overset{\circ}z_0, Z_0,\overset{\circ}Z_0, \overset{\circ}\phi_{\bot0})$ such that the right hand side of  Eq. (\ref{eqhbis}) is positive, the expanding cosmological solutions satisfy $(z(\lambda), Z(\lambda))\to (z_c,Z_c)$ when $\lambda\to +\infty$.
We conclude that the dynamics of the expanding evolutions is attracted by the RDS described by the critical solution $(z, Z)\equiv (z_c,Z_c)$, $\dot \phi_\bot \equiv 0$ \ie $3H^2=c_r^2/a^4$ (Eq. (\ref{propbis})). 


\section{An inflation era ?}
\label{inflation}

We have described how cosmological evolutions can be determined by the R-charges of the massless spectrum. In particular, we found that the dynamics  can be attracted to a radiation era, with stabilized complex structures. In standard cosmology, such an era follows a period of inflation introduced  to explain the observed flatness and homogeneity of our universe, or solve the topological defect problem.  It is thus important to analyze if our attraction mechanism admits periods of accelerated cosmology. 

Using the Einstein equations, the inflation condition $\ddot a>0$ or $\dot H+H^2>0$ can be translated into an inequality between the thermal sources $\rho$, $P$ and the scalar kinetic energies (take $\dot Z=0$ for $n=1$),
\be
{1\over 2}\left(P+{1\over 3}\, \rho\right)<-{1\over 3} \left(\dot\phi^2+{1\over 2}\, \dot Z^2+\dot\phi_\bot^2\right)\, .
\ee
Using the variable $\lambda$ and Eq. (\ref{eqh}) (or (\ref{eqhbis})), we obtain
\be
\label{infla}
p-{1\over 6}\, p_z<-2\left(p-{1\over 3}\, p_z\right){1\over {3\over \K}-1}\le 0\, ,
\ee
where there are two conditions in one: First, $p-p_z/6$ has to be negative, and second it has to be negative enough compared to the right hand side that depends on the kinetic energies.

For the classes of models we considered for $n=1$ (see Eq. (\ref{press})) and $n=2$ (see Eq. (\ref{psym})), one can show that a necessary condition  to have $p-p_z/6<0$ is $n_V<0$. We thus carry on our discussion in this case. It then follows that the quantity $p-p_z/3$ in (\ref{infla}) is always positive. Altogether, a necessary and sufficient condition for an accelerated cosmology is: 
\be
\label{infla2}
n_V<0\qquad \and\qquad \K<\I\quad \where\quad \I=-{3p-{1\over 2}\, p_z\over p-{1\over 2}\, p_z}\,.
\ee
Actually, when deriving (\ref{infla2}) from (\ref{infla}), one also finds that $\I\le 3$. However, this is an empty constraint since the function $\I$ satisfies $\I\le 1$. 


\noindent {\large \bf \em For  a susy breaking with n {\bf = 1}}

\begin{figure}[h!]
\vspace{.2cm}
{\Large \it a)}\hspace{8cm}{\Large \it b)}
\begin{center}
\vspace{-0.4cm}
\includegraphics[height=6.2cm]{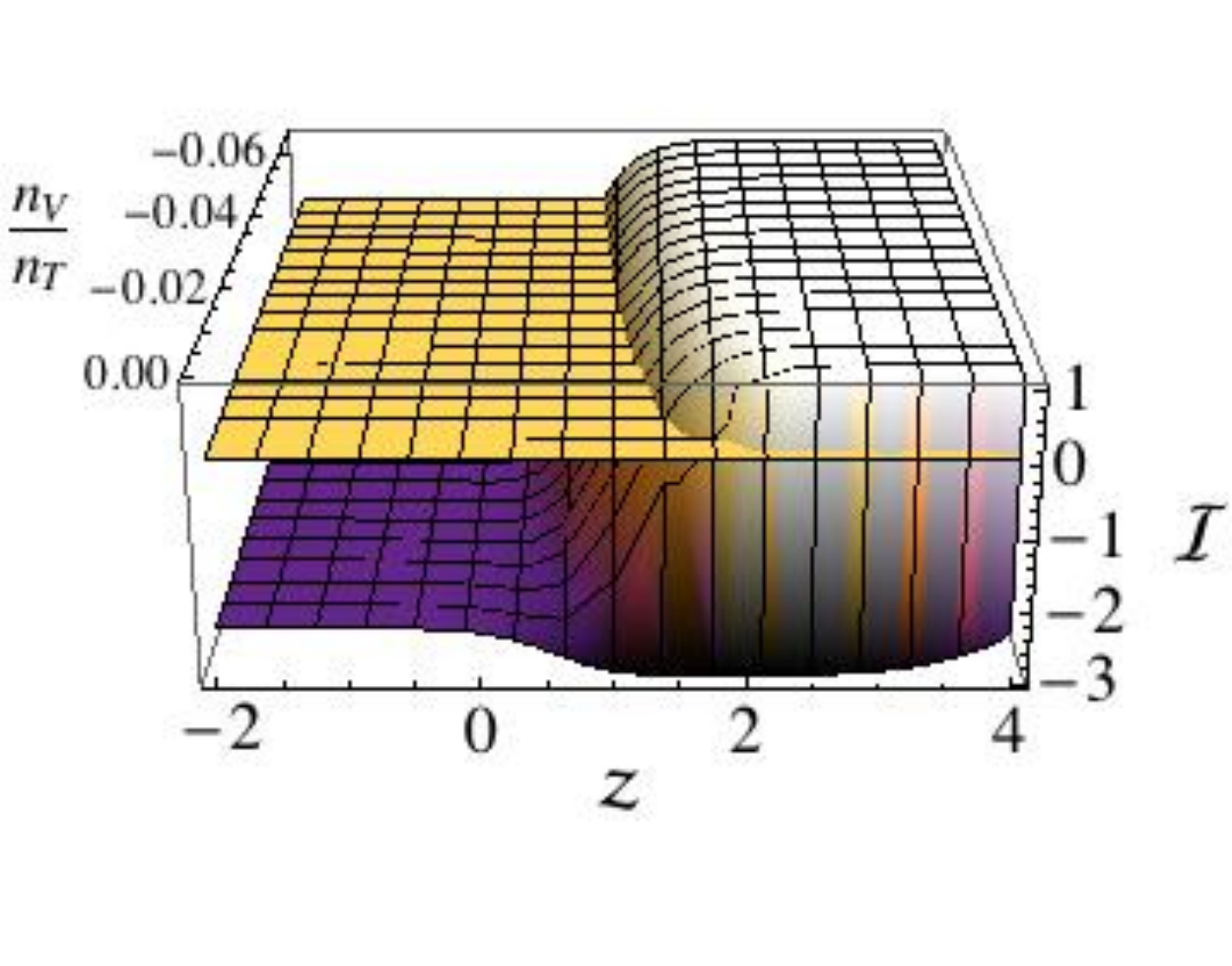}
\includegraphics[height=6.2cm]{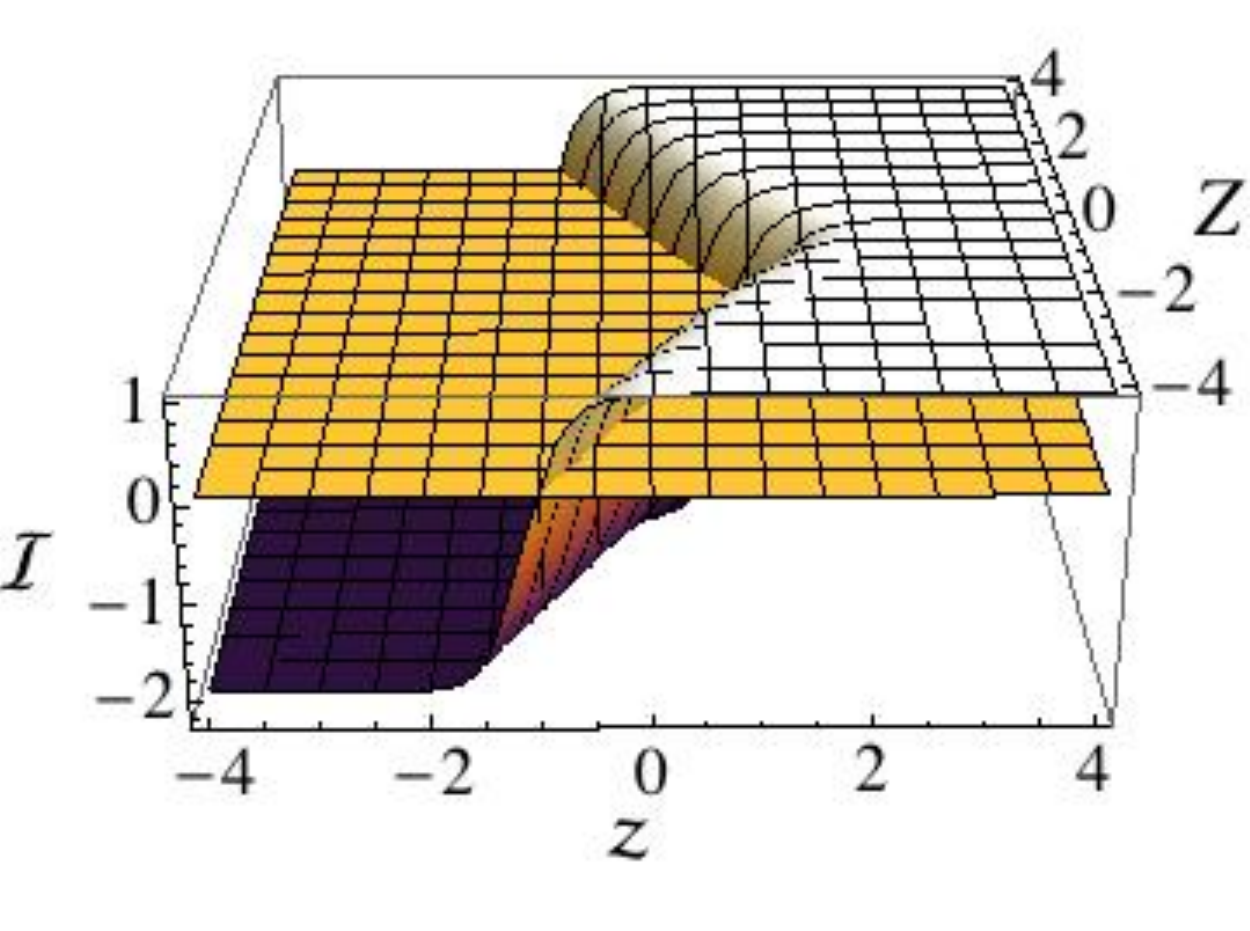}
\vspace{-.5cm}
\caption{\footnotesize \em For one modulus $z$ (Fig. a)): In the domain $\I>0$ of the $(n_V/n_T,z)$-plane ($-1/15<n_V/n_T<0$), $\I$ can be approximated by 1, its value for $z\gg 1$.  For two moduli $(z,Z)$ (Fig. b)): For any fixed $-0.215\lsim n_V/n_T<0$, in the domain $\I>0$ of the $(z,Z)$-plane, $\I$ can be approximated by 1, its asymptotic value in the sectors $II$, $III$, $IV$, $V$ defined in Eq. (\ref{sect}). Fig. b), is drawn for  $n_V/n_T=-1/63$.}  
\vspace{-.4cm}
\label{I}
\end{center}
\end{figure}

\noindent From (\ref{infla2}), in order to have solutions, we need $\I (n_V/n_T,z)>0$. This condition  defines a domain in the $(n_V/n_T,z)$-plan where, by definition, $\I=0$ on the boundary. However, it happens that  $\I\lsim 1$ in the interior, as can be seen on Fig. \ref{I}{\it a}. 
We can thus estimate the $e$-fold number by considering $z$ large and positive. In particular, when $-1/15<n_V/n_T<0$, the $z$'s allowing periods of accelerated cosmology are always larger than $z_c$.  When $z\gg 1$, the condition (\ref{infla2}) simplifies to
\be
\label{ineq}
\K\simeq {1\over 3}\, (\overset{\circ}M/M)^2+(c_\bot/ c_\phi)^2(\overset{\circ}M/M+6)^2 \lsim 1\, ,
\ee
where we have used the second equation of (\ref{eqq}). An era of acceleration exists if $(c_\bot/c_\phi)^2<1/33$. It begins and ends at most when $\overset{\circ}M/M$ saturates the inequality (\ref{ineq}) \ie equals
\be
\label{be}
l_{b,e}={-18(c_\bot/c_\phi)^2\pm \sqrt{3 [1-33 (c_\bot/c_\phi)^2]}\over 1+3(c_\bot/c_\phi)^2}\, .
\ee
To find the scale factors $a_b$ and $a_e$ when this arises, we integrate the first equation of (\ref{eqq}),
\be
\label{sol<}
\ln \left( {a\over a_0}\right)\equiv \lambda-\lambda_0=\gamma\ln \abs\overset{\circ}M/M+6\abs+\gamma_+\ln\abs\overset{\circ}M/M-l_+\abs +\gamma_-\ln\abs\overset{\circ}M/M-l_-\abs\, , 
\ee
where
\be
\begin{array}{c}
\displaystyle l_\pm={-18 (c_\bot/c_\phi)^2\pm 3\sqrt{1-9 (c_\bot/c_\phi)^2}\over 1+3(c_\bot/c_\phi)^2}\, ,
\\
\displaystyle\gamma={3\over 1+3 (c_\bot/c_\phi)^2}\, {1\over (l_++6)(l_-+6)}\; , \quad \gamma_\pm=\pm{3\over 1+3 (c_\bot/c_\phi)^2} \, {1\over (l_\pm+6)(l_+-l_-)}\, .
\end{array}
\ee
Inserting the particular values (\ref{be}) in (\ref{sol<}), the maximum $e$-fold number is found to be
\be
\label{e1m}
e(c_\bot/c_\phi)=\ln{a_e\over a_b}=\gamma\ln \left\abs {l_e+6\over l_b+6}\right\abs+\gamma_+\ln\left\abs {l_e-l_+\over l_b-l_+}\right\abs +\gamma_-\ln\left\abs {l_e-l_-\over l_b-l_-}\right\abs\, , 
\ee
that satisfies $e(c_\bot/c_\phi)\leq e(0)\simeq 0.227$. Thus, even if periods of accelerated scale factor are allowed, they do not give rise to inflationary eras.  


\noindent {\large \bf \em For  a susy breaking with n {\bf = 2}}

\noindent For any fixed $n_V/n_T<0$, the condition $\I (n_V/n_T,z,Z)> 0$ defines a domain in the $(z,Z)$-plan, as shown on Fig. \ref{I}{\it b}. Using the notations of \cite{Cosmo-2}, this zone spans four asymptotic sectors defined as follows,
\be
\label{sect}
\begin{array}{clcl}
II & :~\; Z\to -\infty , ~z\sim \eta Z, ~\abs \eta\abs <{1\over 2}\; , &III  & :~\; Z\sim \eta z\to -\infty , \; -2<\eta<0\; ,\\
IV  &: ~\; Z\sim \eta z\to +\infty , ~0<\eta<2\; , &V & :~\; Z\to +\infty , ~z\sim \eta Z, ~\abs \eta\abs <{1\over 2}\; , \end{array}
\ee
inside of which $\I\lsim 1$. We will evaluate the $e$-fold number by approximating $p(z,Z)$ by its asymptotic expansions, which are well defined in each sectors \cite{Cosmo-2}:
\be
\begin{array}{ll}
p^{II}&\!\!\!\!=e^{4z}e^{-3Z}n_VS^o_6+e^{-z}e^{-Z/2}(n_TS_5^o+n_VS^e_5)+e^{3z}e^{3Z/2}(n_T+n_V)(S^o_4+S^e_4)+\cdots\\
p^{III}&\!\!\!\!=e^{4z}e^{-3Z}n_VS^o_6+e^{4z}e^{2Z}n_V(S_5^o+S^e_5)+2(n_TS_4^o+n_VS_4^e)+\cdots\\
p^{IV}&\!\!\!\!=e^{4z}e^{3Z}n_VS^o_6+e^{4z}e^{-2Z}n_V(S_5^o+S^e_5)+2(n_TS_4^o+n_VS_4^e)+\cdots\\
p^{V}&\!\!\!\!=e^{4z}e^{3Z}n_VS^o_6+e^{-z}e^{Z/2}(n_TS_5^o+n_VS^e_5)+e^{3z}e^{-3Z/2}(n_T+n_V)(S^o_4+S^e_4)+\cdots
\end{array}
\ee
where the dots stand for exponentially suppressed terms and the coefficients $S_{4,5,6}^{o,e}$ are constants\footnote{Their precise values are not needed in the following but can be found in the Appendix of \cite{Cosmo-2}.}. 

It is interesting to note that the behavior of the temperature depends drastically on the location of the representative point of the system in the $(z,Z)$-plan. In sectors $III$ and $IV$, one finds $\A\simeq 1$, $\B\simeq 0$ as in the 1 modulus case, while in sectors $II$ (or $V$), one has $\A\simeq 4/3$, $\B\simeq 1/6$ (or $-1/6$). Using the energy-momentum conservation in the form of Eq. (\ref{phirondbis}), this reaches
\be
II~:~~{M^2\, e^Z\over a^6\,T^8 }=\mbox{cst.}\; , \qquad III \cup IV ~:~~a\, T=\mbox{cst.}\; ,\qquad V~:~~{M^2\, e^{-Z}\over a^6\,T^8 }=\mbox{cst.}\, .
\ee
However, the Eqs of motion for the scalar fields are identical in all these asymptotic sectors,
\be
\label{sys+infbis}
\left\{
\begin{array}{l}
\displaystyle
{1\over {1\over 2}(\overset{\circ}{M}/M)^2+{1\over 4}\, \dot Z^2+{1\over 2}\, \overset{\circ}{\phi}_\bot\!\!\!{}^2-3}(\overset{\circ}{M}/M)^\circ -\overset{\circ}{M}/M-4=0\\ 
\displaystyle{1\over {1\over 2}(\overset{\circ}{M}/M)^2+{1\over 4}\, \dot Z^2+{1\over 2}\, \overset{\circ}{\phi}_\bot\!\!\!{}^2-3}\overset{\circ\circ}Z-\overset{\circ}Z+6=0\\
\displaystyle{1\over {1\over 2}(\overset{\circ}{M}/M)^2+{1\over 4}\, \dot Z^2+{1\over 2}\, \overset{\circ}{\phi}_\bot\!\!\!{}^2-3}\overset{\circ\circ}{\phi}_\bot -\overset{\circ}{\phi}_\bot=0
\end{array}
\right.
\ee
where the fractions in front of the second derivatives must have the sign of $n_V$ \ie negative. When $\overset{\circ}{M}/M+4\neq 0$, it is easy to show that there exist constants $c_\phi\neq 0$, $\cz$, $c_\bot$ such that
\be
\label{eqqbis}
\left\{
\begin{array}{l}
\displaystyle
2(\overset{\circ}{M}/M)^\circ=(\overset{\circ}{M}/ M+4) \P(\overset{\circ}{M}/M) \\ 
\displaystyle\overset{\circ}Z-6=2\, {\cz\over c_\phi}\, ( \overset{\circ}{M}/M+4)\\
\displaystyle\overset{\circ}\phi_\bot=\sqrt{2}\, {c_\bot\over c_\phi}\, ( \overset{\circ}{M}/M+4)
\end{array}
\right.
\ee
where 
\be
\label{Pbis}
\begin{array}{ll}
 \P(x)=&\!\!\![1+2(\cz/ c_\phi)^2+2(c_\bot/ c_\phi)^2]\, x^2+4[4(\cz/c_\phi)^2+3\cz/c_\phi +4 (c_\bot/ c_\phi)^2]\, x\\ \\ 
 &\!\!\!+4 [8(\cz/c_\phi)^2+12\cz/c_\phi +8 (c_\bot/ c_\phi)^2+3 ]~~~~~~~~\mbox{ is of the sign of $n_V$.}
\end{array}
\ee
The condition (\ref{infla2}) is also identical in all asymptotic sectors $II$,...,$V$,
\be
\label{ineqbis}
\K\simeq {1\over 2}\, (\overset{\circ}M/M)^2+[(\cz/ c_\phi)(\overset{\circ}M/M)+3]^2+(c_\bot/ c_\phi)^2(\overset{\circ}M/M+4)^2 \lsim1\, ,
\ee
where we have used the second and third  equations of (\ref{eqqbis}). A period of accelerated cosmology exists if
\be
(c_\bot/c_\phi)^2 <{1\over 14} \; , \qquad c_-<\cz/c_\phi<c_+~~\where~~ c_\pm={-6\pm 2\sqrt{2[1-14(c_\bot/c_\phi)^2]}\over 7}\, .
\ee
It begins and ends at most when $\overset{\circ}M/M$ saturates the inequality (\ref{ineqbis}) \ie equals
\be
\label{bebis}
l_{b,e}={-2 [4(\cz/c_\phi)^2+3\cz/c_\phi+4(c_\bot/c_\phi)^2]\pm \sqrt{14(c_+-\cz/c_\phi)(\cz/c_\phi-c_-)}\over 1+2(\cz/c_\phi)^2+2 (c_\bot/c_\phi)^2}\, .
\ee
To find the corresponding scale factors $a_b$, $a_e$, we integrate the first equation  of (\ref{eqqbis}),
\be
\label{sol<bis}
\ln \left( {a\over a_0}\right)\equiv \lambda-\lambda_0=\gamma\ln \abs\overset{\circ}M/M+4\abs+\gamma_+\ln\abs\overset{\circ}M/M-l_+\abs +\gamma_-\ln\abs\overset{\circ}M/M-l_-\abs\, , 
\ee
where
\be
\begin{array}{l}
\displaystyle l_\pm={-2 [4(\cz/c_\phi)^2+3\cz/c_\phi+4(c_\bot/c_\phi)^2]\pm \sqrt{10(C_+-\cz/c_\phi)(\cz/c_\phi-C_-)}\over 1+2(\cz/c_\phi)^2+2 (c_\bot/c_\phi)^2}\, ,
\\
\displaystyle \qquad \mbox{with} \quad C_\pm={-6\pm \sqrt{7[3-10(c_\bot/c_\phi)^2]}\over 5}\, ,
\\
\displaystyle\gamma={2\over 1+2(\cz/c_\phi)^2+2 (c_\bot/c_\phi)^2}\, {1\over (l_++4)(l_-+4)}\, ,
\\
\displaystyle\gamma_\pm=\pm{2\over 1+2(\cz/c_\phi)^2+2 (c_\bot/c_\phi)^2} \, {1\over (l_\pm+4)(l_+-l_-)}\, .
\end{array}
\ee
Inserting (\ref{bebis}) in (\ref{sol<bis}), the maximum $e$-fold number is found to be
\be
\label{e1m2}
e(\cz/c_\phi,c_\bot/c_\phi)=\ln{a_e\over a_b}=\gamma\ln \left\abs {l_e+4\over l_b+4}\right\abs+\gamma_+\ln\left\abs {l_e-l_+\over l_b-l_+}\right\abs +\gamma_-\ln\left\abs {l_e-l_-\over l_b-l_-}\right\abs\, , 
\ee
that satisfies $e(\cz/c_\phi,c_\bot/c_\phi)\leq e(-3/2,0)\simeq 0.189$. The order of magnitude of the $e$-fold number for $n=1$ and $n=2$ is thus the same. They cannot account for the astrophysical observations. 

\section{Summary, limitations and perspectives}

The present work focuses on an intermediate cosmological era, $t_E\leq t\leq t_w$. The time $t_E$ corresponds to the exit of an Hagedorn phase \ie after the notion of topology and geometry emerge, with 3 large space-like dimensions and possibly internal directions of intermediate scales. The characteristic time $t_w$ is associated to the electroweak phase transition, occurring before large scale structure formations in the universe.

To describe the intermediate era, we consider flat classical 4-dimensional space-times within the context of superstring theory, where $\N=8, 4, 2$ supersymmetry is spontaneously broken. Our aim is to  determine the different cosmological behaviors that can be induced when three kinds of effects are taken into account. Namely,\\
\indent - statistical physics effects at temperature $T$, \\
\indent - radiative corrections of magnitude $M$, the spontaneous supersymmetry breaking scale, \\
\indent - the presence of moduli fields, whose kinetic energies scale as  $1/a^6$. \\
We determine the low energy effective action at the one-loop level and parametrize data at the exit $t_E$ of the Hagedorn phase by a set of IBC.

To be specific, when $n=1$ internal radius is involved in the supersymmetry breaking, we show that, depending on the R-symmetry charges of the massless spectra, the models can be grouped in different {\em Cases (a)}, {\em (b)} or {\em (c)}. For each, there exists a partition of the set of IBC, whose classes yield cosmological evolutions that share a common behavior.  
\begin{itemize}
\item In {\em Case (a)}, one class of IBC seems to induce a spontaneous decompactification of the internal direction involved in the breaking of supersymmetry. It is out of the scope of the present paper.
\item In {\em Case (b)}, the evolutions of the scale factor are monotonic. The expanding solutions belong to the same class and  are attracted to a radiation dominated era. Asymptotically, one has $3H^2\propto 1/a^4\propto T^4\propto M^4$, where $a(t)\propto \sqrt{t}$.  
\item In {\em Case (c)}, a class of IBC reaches solutions attracted to a Big Crunch era dominated by the radiative corrections to the initially vanishing vacuum energy. Asymptotically, one has $3H^2\propto M^4\propto 1/a^{24}\propto T^{24}$, where $a(t)\propto (t_\BC-t)^{1/12}$.
\item In all {\em Cases}, a class of IBC reaches solutions attracted to a Big Crunch era dominated by the classical moduli kinetic energy. Asymptotically, one has $3H^2\propto 1/a^{6}\propto T^{6}\propto M^{1/(1-3/(8\sigma_-))}$, where $\sigma_-$ is given in 
Eq. (\ref{12b}) and $a(t)\propto (t_\BC-t)^{1/3}$.
\end{itemize}

Our analysis has some limitations. First of all, we have supposed that the radii involved in the spontaneous breaking of supersymmetry and the radius of  the Euclidean time are large enough to avoid Hagedorn like singularities in the partition functions. Since we found Big Crunch behaviors where $T\ll M$ are formally diverging, these evolutions must be restricted to $M$ lower than a maximum Hagedorn bound. It would be interesting to relax this constraint by considering models free of such singularities \cite{GravFluxes}. 

Also, we have studied the thermal and quantum corrections at the one-loop level in flat backgrounds where moduli have some kinetic energy. This means that the induced sources $P$, $\rho$ and their backreactions on the ``cold'' classical backgrounds we start with must be small. Thus, we need afterwords to check the consistency of our approach.\\
\indent - For the RDS, we have
\be
P(t)\propto \rho(t)\propto {1\over a(t)^4}\propto {1\over t^2}\longrightarrow 0\, ,
\ee
which shows its validity. \\
\indent - For the Big Crunch attractor  $a(t)\propto (t_\BC-t)^{1/12}$ of the MDS,  both the thermal/quantum sources and the classical moduli kinetic energy are diverging. Formally, one has
\be
P(t)=-\rho(t)\propto M^4\propto {1\over a(t)^{24}}\propto {1\over (t_\BC-t)^2}~~\gg~~ {1\over a^6}\propto {1\over \sqrt{t_\BC-t}}\, ,
\ee
which shows that this solution has to be restricted to times where $P$ and $\rho$ do not exceed too much the classical kinetic energy.\\
\indent - This limitation does not exist for the Big Crunch behavior  $a(t)\propto (t_\BC-t)^{1/3}$. Even if the thermal/quantum sources and the classical moduli kinetic energy are diverging, one has
\be
P(t)=-\rho(t)\propto M^4\propto {1\over a(t)^{24-9/\sigma_-}}\propto {1\over (t_\BC-t)^{8-3/\sigma_-}}~~\ll~~ {1\over a^6}\propto {1\over (t_\BC-t)^2}\, ,
\ee
since $\sigma_-<1/2$. Thus, the backreaction on the classical background is always small. The dynamical attraction to this solution is only limited by the Hagedorn transitions mentioned previously.\\
\indent To go beyond the constraint that the thermal/quantum corrections must be small, one can try to compute the higher string loop corrections and, if possible, use string-string dualities to find the non-perturbative corrections to the effective action. For the models under consideration in this work, the extended supersymmetry is only broken spontaneously, at the scale $M$ and/or by thermal effects at the scale $T$. This means that the UV behavior of the theories is not affected by these soft breaking  and that the renormalization theorems for extended supersymmetries should persist. To find the cosmological evolutions beyond the first order in perturbation theory, both the thermal effective potential and the corrections to the kinetic terms have to be computed.

The attraction phenomena we found can be generalized to models where $n$ internal radii are involved in the spontaneous breaking of supersymmetry. In particular, we have shown that for $n=2$, the attraction  to the radiation era persists. In the process, the complex structure modulus associated to the ratio of the two radii converges to a constant \ie is dynamically stabilized. 

During the convergence to the radiation dominated era, the ratio $M/T$ reaches its critical value exponentially or with damping oscillations. Periods of accelerated expansions are also allowed. However, the $e$-fold number seems to depend weakly  on $n$ and, at least for $n=1,2$, does not give rise to enough periods of inflation ($e$-fold $\leq 0.2$). Since we have considered models with $\N=8, 4, 2$ initial supersymmetry (\ie before spontaneous breaking), it would be interesting to extend our work to $\N=1$ models and see if the order of magnitude of the $e$-fold is higher. 


\section*{Acknowledgements}

We are grateful to I. Antoniadis, J. Estes and N. Toumbas for useful discussions.
H.P. thanks the Ecole Normale Superieure for hospitality. F.B. thanks M. Petropoulos and I. Florakis for fruitful and stimulating discussions.\\
\noindent This work is partially supported by the ANR (CNRS-USAR) contract 05-BLAN-0079-02. The work of F.B and H.P is also supported by the European ERC Advanced Grant 226371, and CNRS PICS contracts  3747 and 4172.






\end{document}